\documentclass[paper]{JHEP3} 

\usepackage{amssymb}
\usepackage{amsmath}
\usepackage{mathrsfs}
\usepackage{graphics,amscd,amsfonts}
\usepackage{epsfig,multicol,multirow}
\usepackage{subfigure,latexsym}
\usepackage{longtable}

\newcommand{\bbC}{{\mathbb C}}
\newcommand{\bbR}{{\mathbb R}}

\newcommand{\bbZ}{{\mathbb Z}}

\newcommand{\cN}{{\cal N}}

\newcommand{\ket}[1]{|#1\,\rangle}

\newcommand{\ba}{\begin{eqnarray}}
\newcommand{\ea}{\end{eqnarray}}
\newcommand{\beq}{\begin{equation}}
\newcommand{\eeq}{\end{equation}}

\newcommand{\eq} {equation}
\newcommand{\eqa} {eqnarray}
\newcommand{\NN} {\mbox {$\nonumber$}}

\def\mathboxit#1{\vbox{\hrule\hbox{\vrule\kern8pt\vbox{\kern8pt
\hbox{$\displaystyle #1$}\kern8pt}\kern8pt\vrule}\hrule}}

\allowdisplaybreaks
\title{
\huge 3d superconformal indices and isomorphisms of M2-brane theories
}

\author{
\large
Masazumi Honda$^{a,b}$ and Yoshinori Honma$^{b,c}$
\vspace*{0.5cm} \\
\llap{$^a$}Department of Particle and Nuclear Physics,\\
Graduate University for Advanced Studies (SOKENDAI),\\
Tsukuba, Ibaraki 305-0801, Japan\\
\llap{$^b$}High Energy Accelerator Research Organization (KEK),\\
Tsukuba, Ibaraki 305-0801, Japan\\
\llap{$^c$}Harish-Chandra Research Institute,\\
Chhatnag Road, Jhusi, Allahabad 211019, India\\
\vspace*{0.5cm} \\
\email{mhonda@post.kek.jp, yhonma@hri.res.in}
}

\preprint{\footnotesize KEK-TH-1576 \\
\footnotesize HRI/ST/1205}

\abstract{
We test several expected isomorphisms between the $U(N) \times U(N)$ ABJM theory 
and $(SU(N) \times SU(N))/\bbZ_N$ theory including the BLG theory 
by comparing their superconformal indices. 
From moduli space analysis, 
it is expected that this equivalence can hold 
if and only if the rank $N$ and Chern-Simons level $k$ are coprime. 
We also calculate the index of the ABJ theory and investigate 
whether some theories with identical moduli spaces are isomorphic or not. 
}

\begin{document}
\setcounter{footnote}{0}

\section{Introduction}
Low-energy limit of $N$ coincident M2-branes on the orbifold $\bbC^4/\bbZ_k$ is captured 
by the 3d $\cN=6$ superconformal Chern-Simons-matter (ABJM) theory 
with the gauge group $U(N)_k \times U(N)_{-k}$ \cite{Aharony:2008ug} (see also \cite{Aharony:2008gk}).
We can take a large $N$ limit of the ABJM theory by using t'Hooft coupling $\lambda=N/k$ and 
this theory still provides fruitful developments in $AdS_4/CFT_3$ correspondence. 
Meanwhile the BLG theory \cite{Bagger:2007jr,Gustavsson:2007vu} 
based on the Lie 3-algebra $[X^a, X^b, X^c]=f^{abc}_{ \ \ \ d} X^d$ 
can also lead us to another description of multiple M2-branes. 
If we take the structure constant $f^{abcd}$ to be totally anti-symmetric,  
then the BLG theory generically has manifest $\cN=8$ supersymmetry and $SO(8)_R$ R-symmetry. 
In spite of such successful structures,
it is known that the only nontrivial solution for a generalized Jacobi identity
is the $A_4$ algebra defined by $f^{abcd}=\epsilon^{abcd}$ \cite{Papadopoulos:2008sk,Gauntlett:2008uf} 
and the resulting $A_4$ BLG theory can be rewritten as 
the $SU(2) \times SU(2)$ ABJM theory \cite{VanRaamsdonk:2008ft}. 
Actually moduli space analysis of this theory \cite{Lambert:2008et,Distler:2008mk} implies that
the interpretation as two indistinguishable M2-branes on $\bbC^4/\bbZ_k$ can be 
possible only for $k=1$ and $k=2$. 
Therefore the role of the $A_4$ BLG theory with higher $k$ has been somewhat unclear.

In~\cite{Lambert:2010ji}, 
an illuminating answer
has been obtained   
by considering $(SU(2)_k \times SU(2)_{-k})/\bbZ_2$ rather than $SU(2)_k \times SU(2)_{-k}$
as  the correct gauge group. 
The authors
have concluded that there are several isomorphisms
even in the quantum level between
\begin{align}
{\mathboxit{
U(2)_k \times U(2)_{-k}\  \textrm{ABJM and $\bbZ_k$ quotient of}
 \ (SU(2)_k \times SU(2)_{-k})/{\bbZ}_2 \ \textrm{BLG\ theory}
\label{iso_N2Zk}
}}
\end{align}
where $k$ is odd.
For $k=2$,  isomorphism between
\begin{align}
{\mathboxit{
U(2)_2 \times U(2)_{-2}\  \textrm{ABJM and}
 \ SU(2)_2 \times SU(2)_{-2} \ \textrm{BLG\ theory}
\label{iso_N2k2}
}}
\end{align}
has been also conjectured.
As we will see in the next section,
the additional $\bbZ_k$ identification in (\ref{iso_N2Zk}) is 
coming from the $U(1)_B$ baryon symmetry
of $(SU(2) \times SU(2))/\bbZ_2$ theory. 
More generally,
they also
proposed that the conjecture (\ref{iso_N2Zk}) can be extended to arbitrary rank $N$ as
\begin{align}
{\mathboxit{
U(N)_k \times U(N)_{-k}\  \textrm{ABJM and $\bbZ_k$ quotient of}
 \ (SU(N)_k \times SU(N)_{-k})/{\bbZ}_N \ \textrm{theory}
\label{iso_anyN}
}}
\end{align}
where $k$ and $N$ are coprime. 

The conjectures (\ref{iso_N2Zk}) for $k=1$ and (\ref{iso_N2k2}) have been 
already tested
by comparing the superconformal indices \cite{Kinney:2005ej,Romelsberger:2005eg,Bhattacharya:2008zy,Bhattacharya:2008bja}
obtained by applying the localization method \cite{Kim:2009wb} (see also \cite{Imamura:2011su}) and
actually nontrivial coincidences have been observed \cite{Bashkirov:2011pt}.
Furthermore,  it has been also found in~\cite{Bashkirov:2011pt} that 
the superconformal indices of 
\begin{align}
{\mathboxit{
U(3)_2 \times U(2)_{-2} \ \textrm{ABJ theory and} \ (SU(2)_4 \times SU(2)_{-4})/{\bbZ}_2 
\ \textrm{BLG theory}
\label{ABJ_BLG}
}}
\end{align}
agree with each other.
Thus, the nontrivial tests\footnote{
In Appendix~\ref{app:partition}, we provide a further evidence for the conjecture (\ref{iso_N2k2}) 
by calculating the partition function on $S^3$.
} beyond the moduli space analysis have been already performed
for the isomorphisms between the ABJ(M) theories with the nontrivial $\cN=8$ SUSY enhancements
and the corresponding BLG theories\footnote{
One
might be curious about $U(3)_1 \times U(2)_{-1}$ and $U(4)_2 \times U(2)_{-2}$ ABJ theories.
However 
it is widely believed that these theories are dual to 
the $U(2)_1 \times U(2)_{-1}$ and $U(2)_2 \times U(2)_{-2}$ ABJ theories
from parity duality, respectively \cite{Aharony:2008gk}.
For a proof by considering the partition functions on $S^3$, see~\cite{Willett:2011gp,Kapustin:2010mh}.
}. 

In this paper, we test the conjecture  (\ref{iso_N2Zk}) 
for the case without $\cN=8$ SUSY enhancements 
by comparing their superconformal indices.
We also check the conjecture (\ref{iso_anyN}) for $N=3$
and investigate whether extensions of the isomorphisms (\ref{iso_N2k2}) and (\ref{ABJ_BLG}) to higher $k$
are possible or not.
This paper is organized as follows. 
In Section~\ref{sec:mod}, we review the argument of \cite{Lambert:2010ji}
about the isomorphism (\ref{iso_anyN}).
In Section \ref{sec:gno}, we briefly look at the charge quantization condition 
of the $(SU(N)_k \times SU(N)_{-k})/\bbZ_N$ theories.
In section \ref{sec:vs}, we describe our calculation of the superconformal indices. 
In Section \ref{sec:test}, we show our results for the indices and test the conjecture (\ref{iso_anyN}).
In Section \ref{sec:extension}, we investigate a possibility 
where the isomorphisms (\ref{iso_N2k2}) and (\ref{ABJ_BLG}) are extended to higher $k$.
Section \ref{sec:con} is devoted to conclusions and discussions.

\section{Dual photon and moduli space of vacua}
\label{sec:mod}
In this section, we review 
arguments of \cite{Lambert:2010ji} about the conjectured isomorphism (\ref{iso_anyN}).
This can be deduced from integrating out the $U(1)$ field or comparing the classical moduli spaces.

The Lagrangian of the $U(N)_k \times U(N)_{-k}$ ABJM theory can 
be expressed \cite{Lambert:2010ji} as
\begin{align}
{\cal{L}}_{u(N) \oplus u(N)}={\cal{L}}_{su(N) \oplus su(N)}^{\textrm{gauged}}+\frac{Nk}{8\pi}
\epsilon^{\mu\nu\lambda}B_{\mu}H_{\nu\lambda},
\end{align}
where $B_{\mu}$ is  the gauge field of the $U(1)_B$ baryon symmetry.
$H_{\mu\nu}$ is the field strength of the trivial $U(1)$, 
which does not couple to all the fields in the $SU(N)_k \times SU(N)_{-k}$ theory.
The second term is the so-called BF term, which is required to make the theory invariant 
under the $\mathcal{N}=6$ supersymmetry 
after gauging the $U(1)_B$ symmetry.
Introducing the Lagrange multiplier $\sigma$ leads to
\begin{align}
{\cal{L}}_{u(N) \oplus u(N)}={\cal{L}}_{su(N) \oplus su(N)}^{\textrm{gauged}}
+\frac{Nk}{8\pi}\epsilon^{\mu\nu\lambda}B_{\mu}H_{\nu\lambda}
+\frac{N}{8\pi} \sigma \epsilon^{\mu\nu\lambda} \partial_\mu H_{\nu\lambda}.
\label{multiplier}
\end{align}
Integrating this by parts, we obtain
\begin{align}
{\cal{L}}_{u(N) \oplus u(N)}={\cal{L}}_{su(N) \oplus su(N)}^{\textrm{gauged}}
+\frac{Nk}{8\pi}\epsilon^{\mu\nu\lambda}B_{\mu}H_{\nu\lambda}
-\frac{N}{8\pi}\epsilon^{\mu\nu\lambda} \partial_{\mu}\sigma H_{\nu\lambda}.
\end{align}
Then, the equation of motion for $H_{\mu\nu}$ is
\begin{align}
B_{\mu}=\frac{1}{k}\partial_{\mu}\sigma.
\end{align}
From this equation, we find
\begin{align}
{\cal{L}}_{u(N) \oplus u(N)}(Z^A, \psi_A, B_{\mu}, H_{\mu\nu\lambda})={\cal{L}}_{su(N) \oplus su(N)}(e^{\frac{i}{k}\sigma}Z^A, 
e^{\frac{i}{k}\sigma}\psi^A) , \ \ \ \ (A=1, \cdots, 4)
\end{align}
where the $U(1)_B$ gauge transformation $B_{\mu} \rightarrow B_{\mu}+\partial_{\mu}\theta$ 
in the language of $\sigma$ is given by 
\begin{align}
\sigma \rightarrow \sigma+k\theta.
\end{align}

The last term of (\ref{multiplier}) implies that
the periodicity of $\sigma$ is determined by the charge quantization condition of $H_{\mu\nu}$.
Note that the charge quantization condition  is different from the usual Dirac quantization condition
since $U(N)$ is not just a product of $U(1)$ and $SU(N)$ 
but  rather it is $(U(1) \times SU(N))/{\mathbb{Z}}_N$.  
Recall that $H$ is a sum of a field strength of each $U(1)$ factor of 
$U(N) \times U(N)$ gauge group.
Finally the condition is given by
\begin{align}
\int dH = \int \frac{1}{2}\epsilon^{\mu\nu\lambda}\partial_{\mu}H_{\nu\lambda} 
\in \frac{4\pi}{N} {\mathbb{Z}} ,
\label{quantization}
\end{align}
which leads the periodicity of $\sigma$ to $2\pi$.
Thus,  we must impose the following identification on the fields
\begin{align}
\hat{Z}^A  \sim e^{\frac{2\pi i}{k}}\hat{Z}^A, \ \ \ \ \ \hat{\psi}^A \sim e^{\frac{2\pi i}{k}}\hat{\psi}^A.
\label{identification}
\end{align}
where we define $\hat{Z}^A$ and $\hat{\psi}_A$ as $\hat{Z}^A=e^{\frac{i}{k}\sigma}Z^A$ and 
$\hat{\psi}_A=e^{\frac{i}{k}\sigma}\psi_A$, respectively.
From this fact, the authors of~\cite{Lambert:2010ji} have concluded that 
the $U(N)_k \times U(N)_{-k}$ ABJM theory is also equivalent to 
a ${\mathbb{Z}}_k$ identification on the $(SU(N)_k \times SU(N)_{-k})/{\mathbb{Z}}_N$ theory. 
As we will see later, this equivalence can hold if we impose an additional constraint on $N$ and $k$.

Next we consider the moduli space of the $(SU(2)_k \times SU(2)_{-k})/{\mathbb{Z}}_2$ theory
with the ${\mathbb{Z}}_k$ identification and check the above result. 
Discussion for generalization to arbitrary rank $N$ is essentially the same \cite{Lambert:2010ji}. 
Setting the scalar potential to be zero,  we can take $Z^A$ up to gauge transformation as
\begin{align}
Z^A=\frac{1}{\sqrt{2}}r^A_1-\frac{i}{\sqrt{2}}r^A_2\sigma_3,
\end{align}
where $r^A_1$ and $r^A_2$ are complex numbers. These can be regarded as the center of mass 
coordinate and the relative coordinate of two M2-branes, respectively. 
For a later convenience, we take
\begin{align}
r^A_1=\frac{1}{2}(z^A_1+z^A_2), \ \ \ \ \  r^A_2=\frac{i}{2}(z^A_1-z^A_2).
\end{align}
Recall that the moduli space of $SU(2) \times SU(2)$ theory is shown to be 
the orbifold $(\bbC^4 \times \bbC^4)/D_{2k}$ \cite{Lambert:2008et,Distler:2008mk}.
Here $D_n$ is the dihedral group of order $2n$,
which is equivalent to the semi-direct product of 
$\bbZ_n$ and $\bbZ_2$ with $\bbZ_2$ acting on $\bbZ_n$ by inversion.
Except for the modification (\ref{quantization}) of the charge quantization condition 
and the ${\mathbb{Z}}_k$ identification (\ref{identification}), 
the same argument holds also in the present case.
Thus, we can show that 
the moduli space of the $(SU(2) \times SU(2))/{\mathbb{Z}}_2$ theory 
with the ${\mathbb{Z}}_k$ identification is given by
\begin{align}
{\mathbb{Z}}_{k}^{U(1)} \ \ \ &: \ \ \ \ \ \ \ z^A_1 \sim e^{\frac{2\pi i}{k}}z^A_1, \ \ \ z^A_2 \sim 
e^{\frac{2\pi i}{k}}z^A_2, \nonumber \\
{\mathbb{Z}}_{2}^{\textrm{perm}} \ \ \ &: \ \ \ \ \ \ \ z^A_1 \sim z^A_2, \nonumber \\
{\mathbb{Z}}_{k}^{(SU(2)\times SU(2))/\bbZ_2}&: \ \ \ \ \ \ \ z^A_1 \sim e^{\frac{2\pi i}{k}}z^A_1, \ \ \ 
z^A_2 \sim e^{-\frac{2\pi i}{k}}z^A_2 ,
\label{su_moduli}
\end{align}
where $\bbZ_{k}^{U(1)}$ is the ${\mathbb{Z}}_k$ identification (\ref{identification}) imposed to $z_1^A ,z_2^A$ 
and $\bbZ^{\textrm{perm}}_2$ is a permutation of two indistinguishable M2-branes. 
Note that the last one is slightly different from the one of the $SU(2) \times SU(2)$ theory given by
\begin{align}
{\mathbb{Z}}_{2k}^{SU(2)\times SU(2)}&: \ \ \ \ \ \ \ z^A_1 \sim e^{\frac{\pi i}{k}}z^A_1, \ \ \ 
z^A_2 \sim e^{-\frac{\pi i}{k}}z^A_2.
\end{align}
The difference comes from  the modified charge quantization condition (\ref{quantization}).
As we mentioned above, $\bbZ^{\textrm{perm}}_2$ and $\bbZ_{k}^{(SU(2)\times SU(2))/\bbZ_2}$ 
identifications generate the dihedral group $D_k$.

The moduli space of $U(2) \times U(2)$ ABJM theory is
\begin{align}
\frac{(\bbR^8/\bbZ_k) \times (\bbR^8/\bbZ_k)}{\bbZ_2},
\end{align}
which is given by the following quotient 
\begin{align}
{\mathbb{Z}}_{k}^{(1)}&: \ \ \ \ \ \ \ z^A_1 \sim e^{\frac{2\pi i}{k}}z^A_1, \ \ \ z^A_2 \sim z^A_2, \nonumber \\
{\mathbb{Z}}_{2}^{\textrm{perm}}&: \ \ \ \ \ \ \ z^A_1 \sim z^A_2, \nonumber \\
{\mathbb{Z}}_{k}^{(2)}&: \ \ \ \ \ \ \ z^A_1 \sim z^A_1, \ \ \ z^A_2 \sim e^{-\frac{2\pi i}{k}}z^A_2 .
\label{u_moduli}
\end{align}
Here $\bbZ_{k}^{(1)}$ and $\bbZ_{k}^{(2)}$ are the $\bbZ_{k}$ identifications of each M2-brane. 
We can easily show that (\ref{su_moduli}) and (\ref{u_moduli}) are equal with each other 
if and only if $k$ is odd\footnote{
If we parametrize $k=2l-1$,  
$\bigl( {\mathbb{Z}}_{k}^{U(1)} \times {\mathbb{Z}}_{k}^{(SU(2)\times SU(2))/\bbZ_2} \bigr)^l$
and $\bigl( ({\mathbb{Z}}_{k}^{U(1)})^{-1} \times {\mathbb{Z}}_{k}^{(SU(2)\times SU(2))/\bbZ_2} \bigr)^l$
are equal to ${\mathbb{Z}}_{k}^{(1)}$ and ${\mathbb{Z}}_{k}^{(2)}$, respectively.
}.
For arbitrary $N$, 
similar discussion leads us to the conclusion 
that the $U(N)_k \times U(N)_{-k}$ ABJM theory is isomorphic to the $\bbZ_k$ quotient of 
the $(SU(N)_k \times SU(N)_{-k})/{\bbZ}_N$ theory if $N$ and $k$ are coprime \cite{Lambert:2010ji} .

\section{Magnetic charge and charge quantization condition}
\label{sec:gno}
Here we briefly look at the charge quantization condition 
of the $(SU(N)_k \times SU(N)_{-k})/\bbZ_N$ theories.
The full global symmetry of the $U(N_1) \times U(N_2)$ ABJ(M) theory is 
$SO(6)_R \times U(1)_T$.
Here $U(1)_T$ is the topological symmetry of the ABJ(M) theory whose conserved current 
is given by\footnote{We use the same notation as in \cite{Bashkirov:2011pt}.}
\begin{align}
J^{\mu}=\frac{1}{16\pi}\epsilon^{\mu\nu\rho}\left( \textrm{Tr}F_{\nu\rho}
+\textrm{Tr}\tilde{F}_{\nu\rho} \right) .
\label{topc}
\end{align}
$F$ and $\tilde{F}$ are the field strengths of $U(N_1)$ and $U(N_2)$ gauge fields, respectively.
The operators carrying the $U(1)_T$ charge are called monopole operators
 \cite{Borokhov:2002ib,Borokhov:2002cg} and involve a non-zero 
magnetic flux in the diagonal $U(1)$ gauge group. 
The monopole operators 
can be labeled by the GNO charges $n_1, \cdots, n_{N_1}$ and $\tilde{n}_1, \cdots, \tilde{n}_{N_2}$ 
which are the monopole charges 
for the Cartan part of the gauge group $U(N_1) \times U(N_2)$ \cite{Goddard:1976qe}. 
The GNO charges label the magnetic flux on $S^2$ surrounding the insertion point of the operator and 
their summation corresponds to the $U(1)_T$ charge as
\begin{align}
Q_T=\frac{k}{4}\left( \sum_{i=1}^{N_1} n_i + \sum_{a=1}^{N_2} \tilde{n}_a \right).
\end{align}
The equations of motion for the gauge fields set 
$\textrm{Tr}F-\textrm{Tr}\tilde{F}=0$ and therefore 
$\sum_i n_i = \sum_a \tilde{n}_a$.
Thus, the $U(1)_T$ charge can be expressed as
\begin{align}
Q_T=\frac{k}{2} T \quad {\rm with}\quad T=  \sum_{i=1}^{N_1} n_i = \sum_{a=1}^{N_2} \tilde{n}_a .
\end{align}

Let us denote $w_i \ (i=1,\cdots, \textrm{rank}(G))$ as the weight vector of the gauge group $G$ 
in an irreducible representation and $\Lambda (G)$ as the weight lattice.
The quantization condition imposes $\exp{(ie\sum_i n_iw_i)}=1$ and this implies that
\begin{align}
e\sum_{i} n_iw_i=2\pi \bbZ
\end{align}
for all $w \in \Lambda(G)$\footnote{This also means that $e\vec{n}$ corresponds to the dual lattice 
$\Lambda^* (G)$. This is a weight lattice of a magnetic (or Langlands) dual group $G^\vee$ and 
$e\vec{n}$ is its weight vector.}.
Here we consider the $SO(3)=SU(2)/\bbZ_2$ and $SU(2)$ as a simple example of dual groups
\footnote{For a general $SU(N)$ group, the dual relation is given by $SU(N)^\vee =SU(N)/\bbZ_N$.}.
In this case, we have
\begin{align}
&\Lambda(SO(3))=\{ 0,\pm 1, \pm2, \pm3,\cdots \}, \nonumber \\
&\Lambda(SU(2))=\{ 0,\pm \frac{1}{2}, \pm1, \pm \frac{3}{2},\cdots \}.
\end{align}
As a result, we find that the magnetic charges must satisfy
\begin{align}
&en_i=2\pi \bbZ \ \ \ \ \ \textrm{for} \ SO(3), \nonumber \\
&en_i=4\pi \bbZ \ \ \ \ \ \textrm{for} \ SU(2).
\end{align}

In the $A_4$ BLG theory, we have $(SU(2) \times SU(2))/\bbZ_2$ gauge group, 
where the $\bbZ_2$ is embedded diagonally in the product of the centers of the two $SU(2)$ factors. 
Note that this is indistinguishable from $(SU(2)/\bbZ_2 )^2$.
Therefore,  
the GNO charges for $(SU(2) \times SU(2))/\bbZ_2$ gauge group 
are allowed to be half the value of those of $SU(2) \times SU(2)$ 
gauge group.
In our notation this implies that
\begin{align}
n_i&=\frac{1}{2} \bbZ \ \ \ \ \ \textrm{for} \ (SU(2) \times SU(2))/\bbZ_2, \nonumber \\
n_i&=\bbZ  \ \ \ \ \ \ \ \textrm{for} \ SU(2) \times SU(2).
\end{align}
A similar discussion can be applied to the $(SU(3) \times SU(3))/\bbZ_3$ theories.
In this case, we finally obtain 
\begin{align}
n_i&=\frac{1}{3} \bbZ \ \ \ \ \ \textrm{for} \ (SU(3) \times SU(3))/\bbZ_3, \nonumber \\
n_i&=\bbZ  \ \ \ \ \ \ \ \textrm{for} \ SU(3) \times SU(3).
\end{align}

As we will see in the next section, 
the superconformal index is given by summation over contributions from each GNO charge. 
Therefore we have to take account of the difference of the charge quantization conditions
for calculating the index.
\section{The superconformal index}
\label{sec:vs}
In this section, 
we derive useful expressions for the superconformal indices of 
the $U(N_1 )_k \times U(N_2 )_{-k}$ ABJ(M) and $SU(N)_k \times SU(N)_{-k}$ theories 
including the BLG theory. 
The superconformal index is defined by
\begin{align}
I(x,z )=\textrm{Tr}\left[ (-1)^F x^{\epsilon+j_3}z^h \right] ,
\end{align}
where $F$, $\epsilon$, $j_3$ and $h$ are 
the fermion number, the energy (or equivalently the conformal dimension),
the projection of spin and the charge of a flavor symmetry, respectively. 
This quantity is a powerful tool for distinguishing theories with a same moduli space.

On the ABJ(M) side, we consider the index with the fixed topological charge $T$.
If there exists an isomorphism as the conjectures (\ref{iso_N2Zk})-(\ref{ABJ_BLG}),
the index of the $SU(N) \times SU(N)$ theory must have contribution 
from charges of certain symmetry corresponding to $U(1)_T$ symmetry.
As noted in \cite{Benna:2008zy}, 
we can write the BLG theory of the product gauge group formulation \cite{VanRaamsdonk:2008ft} 
in $\cN=2$ superspace. 
Of the original $SO(8)_R$ R-symmetry this formulation 
manifestly remains only 
the subgroup $SU(4) \times U(1)_R$.
Because this $U(1)_R$ is not related to the baryonic symmetry,
this has nothing to do with the $U(1)_T$ symmetry in the ABJ(M) theory.
Thus the topological charge $T$ of the ABJ(M) theory 
should correspond to the charge of a $U(1)$ subgroup of the $SU(4) \sim SO(6)$ 
as discussed in \cite{Bashkirov:2011pt}. 
We denote this $U(1)$ subgroup as $U(1)_t$. 
On the BLG side, we treat this $U(1)_t$ as the flavor symmetry
whose charge assignments are $+1(-1)$ to the (anti-)bi-fundamental.
Therefore we introduce the variable $z$ to distinguish the $U(1)_t$ symmetry of the 
$SU(N) \times SU(N)$ theory and compare the index with the one of the ABJ(M) theory.

\subsection{$U(N_1 )_k \times U(N_2 )_{-k}$ ABJ theory}
By applying the localization method \cite{Kim:2009wb,Imamura:2011su}, 
the (whole) superconfomal index of the $U(N_1 )_k \times U(N_2 )_{-k}$ ABJ theory with $z=1$
can be represented as
\begin{\eq}
I_{\rm ABJ} (x ) = \sum_{\{ n\} ,\{ \tilde{n}\}} \frac{x^{\epsilon_0}  }{(\rm sym )} 
        \int_{-\pi}^\pi \frac{d^{N_1} \lambda}{(2\pi )^{N_1}} \frac{d^{N_2} \tilde{\lambda}}{(2\pi )^{N_2}}
          e^{S_0} 
        \exp{\Bigl[ \sum_{p=1}^\infty \frac{1}{p} f_{\rm tot}(x^p ,e^{ip\lambda} ,e^{ip\tilde{\lambda}} ) \Bigr]} ,
\end{\eq}
where $n_i$ and $\tilde{n}_a$ are the GNO charges, $\lambda$ and $\tilde{\lambda}$ are constant 
holonomy zero modes and
\begin{\eqa}
f_{\rm tot} &=& f_{\rm vec} +f_{\rm hyper} ,\NN \\
f_{\rm vec} (x,e^{i\lambda} ,e^{i\tilde{\lambda}} )
&=& -\sum_{i\neq j}^{N_1} \left( e^{i(\lambda_i -\lambda_j )} x^{|n_i -n_j |} \right)
    -\sum_{a\neq b}^{N_2} \left( e^{i(\tilde{\lambda}_a -\tilde{\lambda}_b )} x^{|\tilde{n}_a -\tilde{n}_b |} \right) ,\NN \\
f_{\rm hyper} (x ,e^{i\lambda} ,e^{i\tilde{\lambda}} )
&=& 2\sum_{i,b} \left( 
      \frac{x^{1/2}}{1+x} x^{|n_i -\tilde{n}_b |}  e^{ i(\lambda_i -\tilde{\lambda}_b )}
     +\frac{x^{1/2}}{1+x} x^{|n_i -\tilde{n}_b |}  e^{-i(\lambda_i -\tilde{\lambda}_b )} \right) ,\NN \\
S_0 &=& ik \sum_{i}n_i \lambda_i -ik\sum_a \tilde{n}_a \tilde{\lambda}_a ,\NN \\
\epsilon_0 &=& \sum_{i,b} |n_i -\tilde{n}_b | -\frac{1}{2}\sum_{i,j}|n_i -n_j |     
                                              -\frac{1}{2}\sum_{a,b}|\tilde{n}_a -\tilde{n}_b | . 
\end{\eqa}
The factor ``$({\rm sym})$`` denotes the rank of Weyl group for unbroken gauge group.
By using the formula
\[
\sum_{p=1}^\infty \frac{1}{p} e^{ip \lambda} z^p = -\log{(1-z e^{i\lambda}  )},
\]
the contribution from the vector multiplet is rewritten as
\begin{\eqa}
\exp{\Bigl[ \sum_{p=1}^\infty \frac{1}{p} f_{\rm vec}(x^p ,e^{ip\lambda} ,e^{ip\tilde{\lambda}} ) \Bigr]} 
=\prod_{i\neq j} \left( 1-x^{|n_i -n_j |} e^{i(\lambda_i -\lambda_j )}  \right)
  \prod_{a\neq b} \left( 1-x^{|\tilde{n}_a -\tilde{n}_b |} e^{i(\tilde{\lambda}_a -\tilde{\lambda}_b )}  \right) .     \NN\\
\end{\eqa}
The contribution from the hyper multiplet is a bit more complicated. 
By using
\begin{\eqa}
\sum_{p=1}^\infty \frac{1}{p} \frac{y^p}{1+x^p}  e^{ip \lambda}
&=& -\sum_{m=0}^\infty \log{\left( 1-yx^{2m} e^{i \lambda} \right)}  
    +\sum_{m=0}^\infty \log{\left( 1-yx^{2m+1} e^{i \lambda} \right)} ,  
\end{\eqa}
we obtain
\begin{\eqa}
&& \exp{\Bigl[ \sum_{p=1}^\infty \frac{1}{p} f_{\rm hyper}(x^p ,e^{ip\lambda} ,e^{ip\tilde{\lambda}} ) \Bigr]}  \NN \\
&=& \prod_{i,b}\Biggl[ \prod_{m=0}^\infty 
                \frac{1- x^{2m+3/2 +|n_i -\tilde{n}_b | }  e^{i(\lambda_i -\tilde{\lambda}_b )}}
                     {1- x^{2m+1/2 +|n_i -\tilde{n}_b |}  e^{i (\lambda_i -\tilde{\lambda}_b )}}
                \frac{1- x^{2m+3/2 +|n_i -\tilde{n}_b | }  e^{-i(\lambda_i -\tilde{\lambda}_b )}}
                     {1- x^{2m+1/2 +|n_i -\tilde{n}_b |}  e^{-i (\lambda_i -\tilde{\lambda}_b )}}
               \Biggr]^2 \NN \\
&=& \prod_{i,b}\Biggl[ 
     \frac{(x^{3/2 +|n_i -\tilde{n}_b | }  e^{i(\lambda_i -\tilde{\lambda}_b )};x^2)_\infty}
          {(x^{1/2 +|n_i -\tilde{n}_b | }  e^{i(\lambda_i -\tilde{\lambda}_b)};x^2)_\infty} 
     \frac{(x^{3/2 +|n_i -\tilde{n}_b | }  e^{-i(\lambda_i -\tilde{\lambda}_b )};x^2)_\infty}
          {(x^{1/2 +|n_i -\tilde{n}_b | }  e^{-i(\lambda_i -\tilde{\lambda}_b )};x^2)_\infty} \Biggr]^2 \NN\\
&=& \prod_{i,b}  \mathcal{I}_{\rm hyper} \left( x,n_i -\tilde{n}_b  , e^{i(\lambda_i -\tilde{\lambda}_b)} \right)  ,
\end{\eqa}
where $(a;q)_\infty =\prod_{m=0}^\infty (1-aq^m )$ is the q-Pochhammer symbol and
\begin{\eq}
\mathcal{I}_{\rm hyper} (x,n,y)
= \Biggl[ 
     \frac{(x^{3/2 +|n | } y  ;x^2)_\infty} {(x^{1/2 +|n | }    y ;x^2)_\infty} 
     \frac{(x^{3/2 +|n | }  y^{-1} ;x^2)_\infty} {(x^{1/2 +|n | }    y^{-1} ;x^2)_\infty} \Biggr]^2 .
\end{\eq}
Thus, the superconformal index becomes the following simple form
\begin{\eqa}
I_{\rm ABJ} (x ) 
&=& \sum_{\{ n\} ,\{ \tilde{n}\}} \frac{x^{\epsilon_0}  }{(\rm sym )} 
      \int_{-\pi}^\pi \frac{d^{N_1} \lambda}{(2\pi )^{N_1}} \frac{d^{N_2} \tilde{\lambda}}{(2\pi )^{N_2}} e^{S_0} \NN \\
&&\times  \prod_{i\neq j} \left( 1-x^{|n_i -n_j |} e^{i(\lambda_i -\lambda_j )}  \right)
          \prod_{a\neq b} \left( 1-x^{|\tilde{n}_a -\tilde{n}_b |} e^{i(\tilde{\lambda}_a -\tilde{\lambda}_b )}  \right)
  \NN \\
&&\times \prod_{i,b} \mathcal{I}_{\rm hyper} \left( x,n_i -\tilde{n}_b  , e^{i(\lambda_i -\tilde{\lambda}_b)} \right) .
\end{\eqa}
Next we perform a change of variables as
\begin{\eq}
\mu_i =\lambda_i - \lambda_{N_1} \ (i=1,\cdots ,N_1 -1 ),\quad \mu_{N_1}=\lambda_{N_1},\quad
\nu_a =\tilde{\lambda}_a - \lambda_{N_1} .
\end{\eq}
Then we can easily integrate over $\mu_{N_1}$ and obtain
\begin{\eqa}
I_{\rm ABJ} (x ) 
&=&  \sum_{n_1 ,\cdots ,n_{N_1 -1},\tilde{n}_1 ,\cdots ,\tilde{n}_{N_2}} 
    \frac{x^{\epsilon_0^\prime}  }{(\rm sym )} 
        \int \frac{d^{N_1 -1} \mu}{(2\pi )^{N_1 -1}} \frac{d^{N_2} \nu}{(2\pi )^{N_2}}  e^{S_0^\prime} \NN \\
&&\times  \prod_{i\neq j} \left( 1-x^{|n_i -n_j |} e^{i(\mu_i -\mu_j )}  \right)
          \prod_{a\neq b} \left( 1-x^{|\tilde{n}_a -\tilde{n}_b |} e^{i(\nu_a -\nu_b )}  \right)  \NN \\
&&\times  \prod_{i} \left( 1-x^{|n_i +\sum_j n_j -\sum_a \tilde{n}_a |} e^{i\mu_i }  \right)
                    \left( 1-x^{|n_i +\sum_j n_j -\sum_a \tilde{n}_a |} e^{-i\mu_i }  \right)  \NN \\
&&\times \prod_{i,b}
     \mathcal{I}_{\rm hyper} \left( x,n_i -\tilde{n}_b  , e^{i(\mu_i -\nu_b)} \right) 
    \prod_{a} \mathcal{I}_{\rm hyper} \left( x,\sum_b \tilde{n}_b -\tilde{n}_a -\sum_i n_i  , e^{i \nu_a } \right) ,
     \NN \\
\end{\eqa}
where
\begin{\eqa}
S_0^\prime 
&=& ik \sum_{i=1}^{N_1 -1} n_i \mu_i -ik\sum_a \tilde{n}_a \nu_a , \NN\\
\epsilon_0^\prime 
&=&  \sum_{i,b} |n_i -\tilde{n}_b | -\sum_{i<j}|n_i -n_j |  -\sum_{a<b}|\tilde{n}_a -\tilde{n}_b | \NN \\
&&     +\sum_a \left| -\sum_j n_j +\sum_b \tilde{n}_b -\tilde{n}_a  \right| 
    -\sum_{i}\left| n_i +\sum_j n_j -\sum_b \tilde{n}_b \right| .
\end{\eqa}
Furthermore, we perform a change of the variables as
\begin{\eq}
y_i = e^{i\mu_i},\quad w_a = e^{i\nu_a},
\end{\eq}
where each of them runs over the unit circle in the complex plane.
Then we obtain
\begin{\eqa}
I_{\rm ABJ} (x ) 
&=&  \sum_{n_1 ,\cdots ,n_{N_1 -1},\tilde{n}_1 ,\cdots ,\tilde{n}_{N_2}} 
    \frac{x^{\epsilon_0^\prime}  }{(\rm sym )} 
        \oint \frac{d^{N_1 -1} y}{(2\pi i)^{N_1 -1}} \frac{d^{N_2} w}{(2\pi i )^{N_2}}  
    \prod_i \left( y_i^{kn_i -1} \right) \prod_a \left( w_i^{-k\tilde{n}_a -1} \right) \NN \\
&&\times  \prod_{i\neq j} \left( 1-x^{|n_i -n_j |} y_i y_j^{-1}   \right)
          \prod_{a\neq b} \left( 1-x^{|\tilde{n}_a -\tilde{n}_b |} w_a w_b^{-1}  \right)  \NN \\
&&\times  \prod_{i} \left( 1-x^{|n_i +\sum_j n_j -\sum_a \tilde{n}_a |} y_i  \right)
                    \left( 1-x^{|n_i +\sum_j n_j -\sum_a \tilde{n}_a |} y_i^{-1}  \right)  \NN \\
&&\times \prod_{i,b}
     \mathcal{I}_{\rm hyper} \left( x,n_i -\tilde{n}_b  , y_i w_b^{-1} \right) 
    \prod_{a} \mathcal{I}_{\rm hyper} \left( x,\sum_b \tilde{n}_b -\tilde{n}_a -\sum_i n_i  ,w_a \right) .
     \NN \\
\end{\eqa}
Thus we can write the index with the fixed topological charge $T$ as
\begin{\eqa}
I_{\rm ABJ}^{(T)} (x ) 
&=&  \sum_{n_1 ,\cdots ,n_{N_1 -1},\tilde{n}_1 ,\cdots ,\tilde{n}_{N_2}} 
   \delta_{\sum_a \tilde{n}_a ,T} 
    \frac{x^{\epsilon_0^\prime}  }{(\rm sym )} \NN \\ 
&&      \oint \frac{d^{N_1 -1} y}{(2\pi i)^{N_1 -1}} \frac{d^{N_2} w}{(2\pi i )^{N_2}}  
    \prod_i \left( y_i^{kn_i -1} \right) \prod_a \left( w_i^{-k\tilde{n}_a -1} \right) \NN \\
&&\times  \prod_{i\neq j} \left( 1-x^{|n_i -n_j |} y_i y_j^{-1}   \right)
          \prod_{a\neq b} \left( 1-x^{|\tilde{n}_a -\tilde{n}_b |} w_a w_b^{-1}  \right)  \NN \\
&&\times  \prod_{i} \left( 1-x^{|n_i +\sum_j n_j -\sum_a \tilde{n}_a |} y_i  \right)
                    \left( 1-x^{|n_i +\sum_j n_j -\sum_a \tilde{n}_a |} y_i^{-1}  \right)  \NN \\
&&\times \prod_{i,b}
     \mathcal{I}_{\rm hyper} \left( x,n_i -\tilde{n}_b  , y_i w_b^{-1} \right) 
    \prod_{a} \mathcal{I}_{\rm hyper} \left( x,\sum_b \tilde{n}_b -\tilde{n}_a -\sum_i n_i  ,w_a \right) .
     \NN \\
\label{ABJ_formula}
\end{\eqa}
The integration can be performed 
by expanding the integrand as power series of $y_i ,w_a$ and picking up the poles at the origin.

\subsection{$SU(N)_k \times SU(N)_{-k}$ theory}
Let us consider  the $SU(N)_k \times SU(N)_{-k}$ theory or the $(SU(N)_k \times SU(N)_{-k})/\bbZ_N$ theory including the BLG theory. 
The difference of global structure of the gauge group only 
affects the value of the GNO charges.
As we mentioned above, 
we treat the $U(1)_t$ symmetry as the flavor symmetry
which assigns the flavor charges $+1$ and $-1$
to the bi-fundamental and anti-bi-fundamental multiplets, respectively.
Then, the superconformal index of the $SU(N)_k \times SU(N)_{-k}$ theory is given by
\begin{\eqa}
&& I_{\rm BLG} (x,z ) \NN \\
&=& \sum_{\{ n\} ,\{ \tilde{n}\}} \delta_{\sum_i n_i ,0} \delta_{\sum_i \tilde{n}_i ,0} 
   \frac{x^{\epsilon_0}  }{(\rm sym )} 
        \int \frac{d^N \lambda}{(2\pi )^N} \frac{d^N \tilde{\lambda}}{(2\pi )^N}
        \delta\left( \sum_i \lambda_i \right) \delta\left( \sum_i \tilde{\lambda}_i \right)  e^{S_0}  \NN \\
&&\times  \prod_{i\neq j}\Bigl[ \left( 1-x^{|n_i -n_j |} e^{i(\lambda_i -\lambda_j )}  \right)
               \left( 1-x^{|\tilde{n}_i -\tilde{n}_j |} e^{i(\tilde{\lambda}_i -\tilde{\lambda}_j )}  \right)  \Bigr] \NN \\
&&\times \prod_{i,j} \mathcal{I}_{\rm hyper} \left( x,n_i -\tilde{n}_j  , z e^{i(\lambda_i -\tilde{\lambda}_j)} \right) 
      \NN\\
&=& \sum_{\{ n\} ,\{ \tilde{n}\}} \delta_{\sum_i n_i ,0} \delta_{\sum_i \tilde{n}_i ,0}
  \frac{x^{\epsilon_0}  }{(\rm sym )} 
        \int \frac{d^{N-1} \lambda}{(2\pi )^{N-1}} \frac{d^{N-1} \tilde{\lambda}}{(2\pi )^{N-1}}  e^{S_0}  \NN \\
&&\times  \prod_{i\neq j}\Bigl[ \left( 1-x^{|n_i -n_j |} e^{i(\lambda_i -\lambda_j )}  \right)
               \left( 1-x^{|\tilde{n}_i -\tilde{n}_j |} e^{i(\tilde{\lambda}_i -\tilde{\lambda}_j )}  \right)  \Bigr] \NN \\
&&\times \left.
     \prod_{i,j} \mathcal{I}_{\rm hyper} \left( x,n_i -\tilde{n}_j  , z e^{i(\lambda_i -\tilde{\lambda}_j)} \right) 
 \right|_{\lambda_N = -\sum_{i=1}^{N-1}\lambda_i ,
            \tilde{\lambda}_N = -\sum_{i=1}^{N-1}\tilde{\lambda}_i}  \NN \\
&=& \sum_{\{ n\} ,\{ \tilde{n}\}} \delta_{\sum_i n_i ,0} \delta_{\sum_i \tilde{n}_i ,0}
  \frac{x^{\epsilon_0}  }{(\rm sym )} 
        \oint \frac{d^{N-1} y}{(2\pi i )^{N-1}} \frac{d^{N-1} w}{(2\pi i )^{N-1}}  
         \prod_{i=1}^{N-1} (y_i w_i )^{-1} e^{S_0}  \NN \\
&&\times  \prod_{i\neq j}\Bigl[ \left( 1-x^{|n_i -n_j |} y_i y_j^{-1}  \right)
               \left( 1-x^{|\tilde{n}_i -\tilde{n}_j |} w_i w_j^{-1}  \right)  \Bigr] \NN \\
&&\times \left.
     \prod_{i,j} \mathcal{I}_{\rm hyper} \left( x,n_i -\tilde{n}_j  , z y_i w_j^{-1} \right) 
 \right|_{ y_N  =\prod_{i=1}^{N-1} y_i^{-1}   w_N =\prod_{i=1}^{N-1} w_i^{-1} }. 
\label{BLG_formula}
\end{\eqa}
Similarly to the ABJ case,
the integration can be also performed 
by expanding the integrand as power series of $y_i ,w_i$ and picking up the poles at the origin.

\section{Test of the conjectured isomorphisms}
\label{sec:test}
In this section,
we present our result of the superconformal indices for the $U(N) \times U(N)$ ABJM theory and 
the $(SU(N) \times SU(N))/\bbZ_N$ theory including the BLG theory.
By using our formula (\ref{ABJ_formula}) and (\ref{BLG_formula}), 
we compute the indices and test the conjectured isomorphisms (\ref{iso_anyN})
for $N=2$ and $N=3$.

\subsection{$N=2$}
Here we consider the $U(2)_k \times U(2)_{-k}$ ABJM theory and the $\bbZ_k$ quotient of the 
$(SU(2)_k \times SU(2)_{-k})/\bbZ_2$ BLG theory. 
As we mentioned in Section 2, we must take $k$ to 
be odd in order to match the moduli spaces of the both theories. 
We compute the superconformal indices up to the fifth orders in $x$. 
\begin{table}[htbp]
\begin{longtable}{|l|l|}
  \hline
GNO charges & Index contribution\\
  \hline 
 $T=0$ & $1+4x+12x^2+24x^3+44x^4$ \\
 \hline
 $\ket{0,0}\ket{0,0}$ & $1+4x+12x^2+8x^3+12x^4$\\
 $\ket{1,-1}\ket{1,-1}$ & $16x^3+32x^4$\\  
  \hline 
  $T=1$ & $4x^{\frac{3}{2}}+20x^{\frac{5}{2}}+22x^{\frac{7}{2}}$\\ 
  \hline
$\ket{1,0}\ket{1,0}$ & $4x^{\frac{3}{2}}+20x^{\frac{5}{2}}+22x^{\frac{7}{2}}$\\
\hline
$T=2$ & $17x^3+48x^4$\\
\hline
$\ket{1,1}\ket{1,1}$ & $10x^3+16x^4$\\
$\ket{2,0}\ket{2,0}$ & $7x^3+32x^4$\\
\hline
\caption{The superconformal index of the $U(2)_3\times U(2)_{-3}$ ABJM theory up to ${\cal{O}}(x^5 )$.
A symbol $\ket{n_1 ,n_2 }\ket{\tilde{n}_1,\tilde{n}_2 }$ denotes 
the contribution from the GNO charges $(n_1 ,n_2 ,\tilde{n}_1,\tilde{n}_2 )$.
$T$ represents the topological charge.}
\label{ABJM_N2k3}
\end{longtable}
\end{table}
\begin{table}[htbp]
\begin{longtable}{|l|l|}
\hline
  GNO charges & Index contribution\\
  \hline 
  $\ket{0,0}\ket{0,0}$ & $1+4x+12x^2+8x^3+12x^4$\\
 & $+z^2(3x+8x^2+12x^3+8x^4)+z^{-2}(3x+8x^2+12x^3+8x^4)$\\
    & $+z^4(6x^2+12x^3+12x^4)+z^{-4}(6x^2+12x^3+12x^4)$\\
   & $+z^6(10x^3+16x^4)+z^{-6}(10x^3+16x^4)+15z^8x^4+15z^{-8}x^4$\\
\hline
 $\ket{1/2 ,-1/2}\ket{1/2, -1/2}$ 
  & $z(6x^{\frac32}+22x^{\frac52}+12x^{\frac72})+z^{-1}(6x^{\frac32}+22x^{\frac52}+12x^{\frac72})$\\  
  & $+z^3(4x^{\frac32}+20x^{\frac52}+22x^{\frac72})+z^{-3}(4x^{\frac32}+20x^{\frac52}+22x^{\frac72})$\\
  & $+z^5(10x^{\frac52}+28x^{\frac72})+z^{-5}(10x^{\frac52}+28x^{\frac72})+18z^7x^{\frac72}+18z^{-7}x^{\frac72}$\\
\hline
$\ket{1,-1}\ket{1,-1}$ & $16x^3+32x^4$\\
 & $+z^2(15x^3+32x^4)+z^{-2}(15x^3+32x^4)$\\
 & $+z^4(12x^3+32x^4)+z^{-4}(12x^3+32x^4)$\\
 & $+z^6(7x^3+32x^4)+z^{-6}(7x^3+32x^4)+16z^8x^4+16z^{-8}x^4$\\
\hline
\caption{The superconformal index of the $(SU(2)_3\times SU(2)_{-3})/{\mathbb{Z}}_2$ BLG theory 
up to ${\cal{O}}(x^5 )$.
If we take the additional $\bbZ_3$ quotient,  
only the terms whose powers of $z$ is multiples of 3 remain.
}
\label{BLGZ2_k3}
\end{longtable}
\end{table}

Let us compare the ABJM index in an individual topological charge $T$ with the BLG index in 
a particular monomial of $z$. 
First, we consider the case for $k=3$.
In Table \ref{ABJM_N2k3}, 
we show the contributions from each GNO charge to the index 
in the $U(2)_3 \times U(2)_{-3}$ ABJM theory.
To summarize the result, the ABJM indices with the fixed topological charge $T$ are given by\footnote{
Here we explicitly show the results only for non-negative $T$
since  $I_{\rm ABJM}^{(-T)}(x) = I_{\rm ABJM}^{(T)}(x)$.
}
\begin{\eqa}
I_{\rm{ABJM},k=3}^{(T=0)}(x) &=& 1+4x+12x^2+24x^3+44x^4 ,\NN\\
I_{\rm{ABJM},k=3}^{(T=1)}(x) &=& 4x^{\frac32}+20x^{\frac52}+22x^{\frac72} ,\NN\\
I_{\rm{ABJM},k=3}^{(T=2)}(x) &=& 17x^3+48x^4 ,
\label{ABJMresult_N2k3}
\end{\eqa}
up to ${\cal{O}}(x^5 )$.
The result of the $(SU(2)_3 \times SU(2)_{-3})/\bbZ_2$ BLG theory is shown in Table \ref{BLGZ2_k3}.
Note that  we have to sum over all relevant GNO charges on the BLG side 
in order to obtain all the contributions to the fixed charge $U(1)_t$. 
The BLG index is summarized as
\begin{align}
I_{\rm{BLG},k=3} (x,z)
&=1+4x+12x^2+24x^3+44x^4 \nonumber \\
& \ \ \ +z(6x^{\frac{3}{2}}+22x^{\frac{5}{2}}+12x^{\frac{7}{2}})+z^{-1}(6x^{\frac{3}{2}}+22x^{\frac{5}{2}}+12x^{\frac{7}{2}}) \nonumber \\
& \ \ \ +z^2(3x+8x^2+27x^3+40x^4)+z^{-2}(3x+8x^2+27x^3+40x^4) \nonumber \\
& \ \ \ +z^3(4x^{\frac32}+20x^{\frac52}+22x^{\frac72})+z^{-3}(4x^{\frac32}+20x^{\frac52}+22x^{\frac72}) \nonumber \\
& \ \ \ +z^4(6x^2+24x^3+44x^4)+z^{-4}(6x^2+24x^3+44x^4) \nonumber \\
& \ \ \ +z^5(10x^{\frac{5}{2}}+28x^{\frac{7}{2}})+z^{-5}(10x^{\frac{5}{2}}+28x^{\frac{7}{2}}) \nonumber \\
& \ \ \ +z^6(17x^3+48x^4)+z^{-6}(17x^3+48x^4) \nonumber \\
& \ \ \ +18z^7x^{\frac{7}{2}}+18z^{-7}x^{\frac{7}{2}}+31z^8x^4+31z^{-8}x^4 ,
\end{align}
up to ${\cal{O}}(x^5 )$.
After taking the additional $\bbZ_3$ quotient,  
several terms are projected out. 
The remaining terms have only specific powers of $z$ which are multiples of 3.
Thus, we obtain the index of the $\bbZ_3$ quotient of 
the $(SU(2)_3 \times SU(2)_{-3})/\bbZ_2$ BLG theory as
\begin{align}
I_{\rm{BLG},k=3}^{\bbZ_3 \ \textrm{quotient}} (x,z)
&=1+4x+12x^2+24x^3+44x^4 \nonumber \\
& \ \ \ +z^3(4x^{\frac32}+20x^{\frac52}+22x^{\frac72})+z^{-3}(4x^{\frac32}+20x^{\frac52}+22x^{\frac72}) \nonumber \\
& \ \ \ +z^6(17x^3+48x^4)+z^{-6}(17x^3+48x^4).
\end{align}
Comparing this with the ABJM indices (\ref{ABJMresult_N2k3}),
the BLG index can be written as 
\begin{\eqa}
I_{\rm{BLG},k=3}^{\bbZ_3 \ \textrm{quotient}} (x,z)
&=& I_{\rm{ABJM},k=3}^{(T=0)}(x)  
     +I_{\rm{ABJM},k=3}^{(T=1)}(x) z^3 +I_{\rm{ABJM},k=3}^{(T=-1)}(x) z^{-3} \NN\\
&&   +I_{\rm{ABJM},k=3}^{(T=2)}(x) z^6 +I_{\rm{ABJM},k=3}^{(T=-2)}(x) z^{-6}  . 
\end{\eqa}
Thus, we find that  
the proposal (\ref{iso_N2Zk}) of \cite{Lambert:2010ji} is correct for $k=3$ at least up to ${\cal{O}}(x^5 )$.

We also show the results for other values of $k$ in Appendix \ref{sec:ABJMN2} and \ref{sec:BLG}.
From Tables \ref{ABJM_N2k5}, \ref{ABJM_N2k7}, \ref{BLGZ2_k5} and \ref{BLGZ2_k7},
we can easily find 
\begin{\eqa}
I_{\rm{BLG},k=5}^{\bbZ_5 \ \textrm{quotient}} (x,z)
&=& 1+4x+12x^2+32x^3 
        +(6x^{\frac{5}{2}}+28x^{\frac{7}{2}} )z^5 +(6x^{\frac{5}{2}}+28x^{\frac{7}{2}} )z^{-5} \NN\\
&=& I_{\rm{ABJM},k=5}^{(T=0)}(x)  
        +I_{\rm{ABJM},k=5}^{(T=1)}(x) z^5 +I_{\rm{ABJM},k=5}^{(T=-1)}(x) z^{-5}  \NN\\
I_{\rm{BLG},k=7}^{\bbZ_7 \ \textrm{quotient}} (x,z)
&=& 1+4x+12x^2+8x^3+12x^4 +8x^{\frac{7}{2}} z^7 +8x^{\frac{7}{2}} z^{-7} \NN\\
&=& I_{\rm{ABJM},k=7}^{(T=0)}(x)  
      +I_{\rm{ABJM},k=7}^{(T=1)}(x) z^7 +I_{\rm{ABJM},k=7}^{(T=-1)}(x) z^{-7} ,
\end{\eqa}
up to ${\cal{O}}(x^5 )$.
Again we can see again that the precise matching is revealed after we impose the additional 
identification $\bbZ_k$ and topological charge $T$ of the ABJM theory has 
the one-to-one correspondence with the $U(1)_t$ charge of the BLG theory.
By contrast, there are no matching for $k=6$ from Tables \ref{ABJM_N2k6} and \ref{BLGZ2_k6}.
These are consistent with the conjecture (\ref{iso_N2Zk}).

\subsection{$N=3$}
As we have seen in Section 2, 
the $U(3)_k \times U(3)_{-k}$ ABJM theory\footnote{
The indices of the $U(3)_k \times U(3)_{-k}$ ABJM theory for $k=1,2$ have been also
calculated in \cite{Gang:2011xp}. See also \cite{Cheon:2012be}.
} would also be isomorphic to the 
$\bbZ_k$ quotient of the $(SU(3)_k \times SU(3)_{-k} )/\bbZ_3$ theory. 
Since the expected isomorphism (\ref{iso_anyN}) can hold iff N and k are coprime, 
$k$ must not be multiples of 3 in this case.

First let us consider the case for $k=1$.
From Table \ref{SU3_k1} in Appendix \ref{sec:SU3}, 
we find the index of the $(SU(3)_1 \times SU(3)_{-1} )/\bbZ_3$ theory as
\begin{\eqa}
I_{SU(3),k=1} (x,z)
&=&   1+8x+71x^2 +320x^3 +(2x^{1/2} +24x^{3/2} +156x^{5/2})z  \NN\\
&&   +(6x +56x^2 +293x^3 ) z^2 +(14x^{3/2} +114x^{5/2})z^3
\end{\eqa}
up to ${\cal{O}}(x^4 )$.
Since the additional $\bbZ_{k=1}$ identification for this case is trivial,
we can easily see from Table \ref{ABJM_N3k1} in Appendix \ref{sec:ABJM_N3} that
\begin{\eqa}
I_{SU(3),k=1}^{\bbZ_1 \ \textrm{quotient}} (x,z)
&=&    I_{\rm{ABJM},k=1}^{(T=0)}(x)  
        +I_{\rm{ABJM},k=1}^{(T=1)}(x) z^1 +I_{\rm{ABJM},k=1}^{(T=-1)}(x) z^{-1}  
        +I_{\rm{ABJM},k=1}^{(T=2)}(x) z^2 \NN\\
&&    +I_{\rm{ABJM},k=1}^{(T=-2)}(x) z^{-2}  
        +I_{\rm{ABJM},k=1}^{(T=3)}(x) z^3 +I_{\rm{ABJM},k=1}^{(T=-3)}(x) z^{-3}  .
\end{\eqa}
The results for other values are also presented in Appendix \ref{sec:SU3} and \ref{sec:ABJM_N3}.
From Tables \ref{ABJM_N3k2}, \ref{ABJM_N3k4}, \ref{ABJM_N3k5}, 
\ref{SU3_k2}, \ref{SU3_k4} and \ref{SU3_k5}, we also find 
\begin{\eqa}
I_{SU(3),k=2}^{\bbZ_2 \ \textrm{quotient}} (x,z)
&=&  1+4x+21x^2 +92x^3 +(3x +16x^2 +87x^3 )z^2 +(3x +16x^2 +87x^3 )z^{-2}   \NN\\
&&  +(11x^2 +60x^3 )z^4 +(11x^2 +60x^3 )z^{-4} \NN\\
&=&    I_{\rm{ABJM},k=2}^{(T=0)}(x)  
        +I_{\rm{ABJM},k=2}^{(T=1)}(x) z^2 +I_{\rm{ABJM},k=2}^{(T=-1)}(x) z^{-2} \NN\\  
&&    +I_{\rm{ABJM},k=2}^{(T=2)}(x) z^4 +I_{\rm{ABJM},k=2}^{(T=-4)}(x) z^{-4} \NN\\
I_{SU(3),k=4}^{\bbZ_4 \ \textrm{quotient}} (x,z)
&=& 1+4x +12x^2 +32x^3 +(5x^2 +24x^3 )z^4 +(5x^2 +24x^3 )z^{-4} \NN\\
&=&    I_{\rm{ABJM},k=4}^{(T=0)}(x)  
        +I_{\rm{ABJM},k=4}^{(T=1)}(x) z^4 +I_{\rm{ABJM},k=4}^{(T=-1)}(x) z^{-4} \NN\\  
I_{SU(3),k=5}^{\bbZ_5 \ \textrm{quotient}} (x,z)
&=& 1+4x +12x^2 +32x^3 +6 x^{5/2} z^5   +6 x^{5/2} z^{-5} \NN\\
&=&    I_{\rm{ABJM},k=5}^{(T=0)}(x)  
        +I_{\rm{ABJM},k=5}^{(T=1)}(x) z^5 +I_{\rm{ABJM},k=5}^{(T=-1)}(x) z^{-5}   
\end{\eqa}
up to ${\cal{O}}(x^4 )$.
Therefore, we find that the isomorphism (\ref{iso_anyN}) for $N=3$ is also correct 
for various values of $k$
at least up to ${\cal{O}}(x^4)$
while there are no matching for $k=3$ from Tables \ref{ABJM_N3k3} and \ref{SU3_k3}.
This is consistent with the conjecture (\ref{iso_anyN}).

\section{Search for 
a possibility of extended isomorphism with higher $k$
}
\label{sec:extension}
In this section, let us consider
whether an extension of the isomorphisms (\ref{iso_N2k2}) and (\ref{ABJ_BLG}) to higher $k$ is
possible or not.
We compute the superconformal indices of theories with an identical moduli space
and compare the results of these theories.

\subsection{
$U(2+l)_k \times U(2)_{-k}$ ABJ theory v.s.
$(SU(2)_{k^2}\times SU(2)_{-k^2} )/\mathbb{Z}_2$ BLG theory 
}
If two theories are isomorphic,
these theories should have a same moduli space.
Since the moduli space of the $(SU(2)_{k^2}\times SU(2)_{-k^2} )/\mathbb{Z}_2$ BLG theory is
same as the one of the $U(2+l)_k \times U(2)_{-k}$ ABJ theory ($0\leq l\leq |k|$),
these pairs would be candidates for the extension of the isomorphism (\ref{ABJ_BLG}) with higher $k$.
Actually if we take $k=2$ and $l=1$, this is nothing but the pair of (\ref{ABJ_BLG}).

First, let us consider the case for $k=3$.
Then, the $(SU(2)_{9}\times SU(2)_{-9} )/\mathbb{Z}_2$ BLG theory has
the same moduli space with
the $U(2)_3 \times U(2)_{-3}$, $U(3)_3 \times U(2)_{-3}$, $U(4)_3 \times U(2)_{-3}$ and
$U(5)_3 \times U(2)_{-3}$ ABJ theories. 
Since it is widely believed that
$U(4)_3 \times U(2)_{-3}$ and $U(5)_3 \times U(2)_{-3}$ ABJ theories are
equivalent to the $U(2)_3 \times U(2)_{-3}$ and $U(3)_3 \times U(2)_{-3}$ theories 
via parity duality \cite{Aharony:2008gk},
we
can concentrate only on the $U(2)_3 \times U(2)_{-3}$ and $U(3)_3 \times U(2)_{-3}$ theories.
From Table \ref{BLG_Z2k9},
we can pick up the result of the $(SU(2)_{9}\times SU(2)_{-9} )/\mathbb{Z}_2$ BLG theory as
\begin{\eqa}
I_{\rm{BLG},k=9} (x,z)
&=&   1 + 4 x + 12 x^2 + 8 x^3 + 12 x^4 +(3x +8x^2 +12x^3 +8x^4 )z^2 \NN\\
&&   +(3x +8x^2 +12x^3 +8x^4 )z^{-2} +(6x^2 +12x^3 +12x^4 )z^4 \NN\\
&&   +(6x^2 +12x^3 +12x^4 )z^{-4} ,
\end{\eqa}
up to ${\cal{O}}(x^5 )$.
On the other hand, the results of
the $U(2)_3 \times U(2)_{-3}$ and $U(3)_3 \times U(2)_{-3}$ ABJ(M) theories 
from Tables \ref{ABJM_N2k3} and \ref{U32_k3} are 
\begin{\eqa}
I_{U(2)\times U(2),k=3}^{(T=0)}(x) &=& 1+4x+12x^2+24x^3+44x^4 ,\NN\\
I_{U(2)\times U(2),k=3}^{(T=1)}(x) &=& 4x^{\frac32}+20x^{\frac52}+22x^{\frac72} ,\NN\\
I_{U(2)\times U(2),k=3}^{(T=2)}(x) &=& 17x^3+48x^4 ,    \\
\NN \\
I_{U(3)\times U(2),k=3}^{(T=0)}(x) &=& 1+4x+12x^2+28x^3+37x^4 ,\NN\\
I_{U(3)\times U(2),k=3}^{(T=1)}(x) &=& 4x^{\frac{3}{2}}+20x^{\frac{5}{2}} +26x^{\frac{7}{2}} ,\NN\\
I_{U(3)\times U(2),k=3}^{(T=2)}(x) &=& 17x^3 +48x^4 . 
\end{\eqa}

We can easily see that the result of the BLG theory does not match with the
calculations of 
the $U(2)_3 \times U(2)_{-3}$ and the $U(3)_3 \times U(2)_{-3}$ ABJ theories.
In particular, there are several terms whose powers of $z$
do not correspond to 
integer $T$ in the ABJ theories. 
Therefore, although the $(SU(2)_{9}\times SU(2)_{-9} )/\mathbb{Z}_2$ BLG theory has 
the same moduli space with the $U(2+l)_3 \times U(2)_{-3}$ ABJ theory,
there are no isomorphisms
among these theories.

Next, let us consider a case for $k=4$.
Then, the $(SU(2)_{16}\times SU(2)_{-16} )/\mathbb{Z}_2$ BLG theory has
the same moduli space with the $U(2+l)_4 \times U(2)_{-4}\ (l=0,1, \cdots ,4)$  ABJ theories. 
Similar to the $k=3$ case, 
we 
can concentrate on the indices
for $l=0,1,2$.
Table \ref{BLG_Z2k16} shows the result of the $(SU(2)_{16}\times SU(2)_{-16} )/\mathbb{Z}_2$ BLG theory.
Again we can easily see that the results do not
agree with each other.
\begin{\eqa}
I_{\rm{BLG},k=16} (x,z)
&=&   1+ 4x  +12x^2 +8x^3 +12x^4 +15x^4 z^8 +15 x^4 z^{-8} +\left(16 x^4+10 x^3 \right) z^6 \NN\\
&& +(16 x^4+10 x^3 )z^{-6} +\left( 12 x^4+12 x^3+6 x^2\right) z^4
       +(12 x^4+12 x^3+6x^2 )z^{-4} \NN\\
&&  +\left( 8 x^4+12 x^3+8 x^2+3 x \right) z^2 +(8 x^4+12 x^3+8 x^2+3 x)z^{-2} ,
\end{\eqa}
up to ${\cal{O}}(x^5 )$.
From Tables \ref{ABJM_N2k4}, \ref{U32_k4} and \ref{U42_k4}, we pick up the results of the ABJ theories as
\begin{\eqa}
I_{U(2)\times U(2),k=4}^{(T=0)}(x) &=& 1+4x+12x^2+8x^3+37x^4 ,\NN\\
I_{U(2)\times U(2),k=4}^{(T=1)}(x) &=& 5x^2 +24x^3 +23x^4 ,\NN\\
I_{U(2)\times U(2),k=4}^{(T=2)}(x) &=& 17x^3+48x^4 , \label{U22_k4} \\
\NN\\
I_{U(3)\times U(2),k=4}^{(T=0)}(x) &=& 1+4x+12x^2+12x^3+30x^4 ,\NN\\
I_{U(3)\times U(2),k=4}^{(T=1)}(x) &=& 5x^2 +24x^3 +28x^4 ,\NN\\
I_{U(3)\times U(2),k=4}^{(T=2)}(x) &=& 24x^4 ,  \label{U32_k4} \\ 
\NN\\
I_{U(4)\times U(2),k=4}^{(T=0)}(x) &=& 1+4x+12x^2+12x^3+31x^4 ,\NN\\
I_{U(4)\times U(2),k=4}^{(T=1)}(x) &=& 5x^2 +24x^3 +28x^4 . \label{U42_k4}
\end{\eqa}
Thus, we conclude that
extension of the isomorphism (\ref{ABJ_BLG}) to higher $k$ 
seems to be impossible.

\subsection{
$U(2+l)_k \times U(2)_{-k}$ ($0\leq l\leq |k|$) ABJ theory v.s.
$SU(2)_{k^2 /2}\times SU(2)_{-k^2 /2} $ BLG theory
}
Similarly we can find the pairs with same moduli spaces:
the $SU(2)_{k^2 /2}\times SU(2)_{-k^2 /2} $ BLG theory and 
the $U(2+l)_k \times U(2)_{-k}$ ($0\leq l\leq |k|$) ABJ theories.
Actually if we set $k=4$ and $l=0$, this becomes the pair of (\ref{iso_N2k2}).
First, let us consider the case for $k=4$.
Then the $SU(2)_{8}\times SU(2)_{-8} $ BLG theory has
a same moduli space with the $U(2+l)_4 \times U(2)_{-4}\ (l=0,1,2,3,4)$  ABJ theories. 
In this case, we can restrict to the cases for $l=0,1,2$ 
as before.
Table \ref{BLG_k8} shows the result of the $SU(2)_{8}\times SU(2)_{-8}$ BLG theory, 
which is given by
\begin{\eqa}
I_{\rm{BLG},k=8} (x,z)
&=&  1+ 4x  +12x^2 +8x^3 +12x^4  \NN\\
&&   +(3x +8x^2 +12x^3 +8x^4 )z^2 +(3x +8x^2 +12x^3 +8x^4 )z^{-2} \NN\\
&&  +(6x^2 +12x^3 +12x^4 )z^4 +(6x^2 +12x^3 +12x^4 )z^{-4} . 
\end{\eqa}
Comparing this with (\ref{U22_k4}), (\ref{U32_k4}) and (\ref{U42_k4}),
we can easily see again that the results do not match with each other.
Thus, we conclude that
extension of the isomorphism (\ref{iso_N2k2})  to higher $k$
might be impossible.

\section{Conclusions}
\label{sec:con}
In this paper we calculated the superconformal indices of the $U(N_1)_k \times U(N_2)_{-k}$ ABJ(M) 
theories and $(SU(N)_k \times SU(N)_{-k})/\bbZ_N$ theories including the BLG theories 
for various values of the rank and the Chern-Simons level. 
We utilize the indices to test 
the conjectured isomorphism between several M2-brane theories
beyond the classical moduli space analysis. 
Actually we have been confirmed the isomorphism between 
\begin{\eqa*}
{\mathboxit{
U(2)_k \times U(2)_{-k}\  \textrm{ABJM and $\bbZ_k$ quotient of}
 \ (SU(2)_k \times SU(2)_{-k})/{\bbZ}_2 \ \textrm{BLG\ theory} 
}}
\end{\eqa*}
for the cases without the $\mathcal{N}=8$ SUSY enhancement.
Since the $(SU(2) \times SU(2))/\bbZ_2$ theory can be expressed by the $A_4$ BLG theory, 
this verification enables us to understand the significance of the $A_4$ BLG theory 
with the higher Chern-Simons level $k>2$. 
By comparing the indices with the fixed topological charge of the ABJM theory
with the contributions from the corresponding charge of the BLG theory, 
we have been obtained the clear understanding for the correspondence. 
We have also tested the conjectured equivalence between 
\begin{align*}
{\mathboxit{
U(3)_k \times U(3)_{-k}\  \textrm{ABJM and $\bbZ_k$ quotient of}
 \ (SU(3)_k \times SU(3)_{-k})/{\bbZ}_3 \ \textrm{BLG\ theory}
}}
\end{align*}
and
it turns out that the isomorphism holds for various values of $k$ at least up to $\mathcal{O}(x^4 )$. 
Moreover we investigated 
a possibility of  extensions of  isomorphisms 
(i)   $U(2)_2 \times U(2)_{-2}$ ABJM and $SU(2)_2 \times SU(2)_{-2} $ BLG theory
and (ii) $U(3)_2 \times U(2)_{-2}$ ABJ theory and $(SU(2)_4 \times SU(2)_{-4})/{\bbZ}_2$ BLG theory
to higher $k$.
Comparing the indices of theories with an identical moduli space,
we have found that such extensions might be impossible.

Important subject related to our work is concerned 
with the $-3\lambda^2/8N^2$ discrepancy in the $AdS_4/CFT_3$ correspondence.
In \cite{Fuji:2011km}, apart from the worldsheet instanton contributions and the constant map contributions, the all 
genus free energy of the ABJM matrix model was resummed to
the Airy function 
which depends on the ``renormalized" t'Hooft coupling $\lambda_{\textrm{ren}}$ given by
\begin{align}
\lambda_{\textrm{ren}}=\lambda-\frac{1}{24}-\frac{\lambda^2}{3N^2}.
\label{lamren}
\end{align}
This shift was originally observed in \cite{Drukker:2011zy} by simplifying the expression of the all genus free energy.
Note that this renormalization is consistent with the Fermi gas approach \cite{Marino:2011eh}, 
numerical calculation \cite{Hanada:2012si} 
and exact calculation for $k=1$ \cite{Hatsuda:2012hm,Putrov:2012zi}.
However, this renormalization of the t'Hooft coupling is slightly different from the expectation from the 
gravity side \cite{Bergman:2009zh}:
\begin{align}
\lambda_{\textrm{ren}}^{\textrm{grav}}=\lambda-\frac{1}{24}+\frac{\lambda^2}{24N^2}.
\label{lamrengra}
\end{align}
This shift comes from the higher curvature correction $C_3 \wedge I_8$ in M-theory. Here $I_8$ is a 
8-form anomaly polynomial \cite{Duff:1995wd}. Although (\ref{lamren}) and (\ref{lamrengra}) agree in 
the large $N$ limit, there is a discrepancy $-3\lambda^2/8N^2$ at the non-planar level.
From the aspect of testing $AdS_4/CFT_3$ duality in quantum level, 
we should definitely obtain more understanding on both the gauge theory side and the gravity side.
As discussed in \cite{Fuji:2011km}, 
a possible resolution on the matrix model side is to consider the effect of $U(1)$ 
factors in the gauge group $U(N) \times U(N)$ which provide finite $N$ correction. 
Although the current status of this problem is unclear, it is worth revisiting the $U(1)$ factors in ABJM theory in greater detail. 

Recently there have been some arguments about applying the Lie 3-algebra to the M5-branes 
\cite{Lambert:2010wm} 
(see also \cite{Honma:2011br,Kawamoto:2011ab}). Although the significance of the Lie 
3-algebra in 6d $\cN=(2,0)$ theory is not so clear, it is valuable to keep in mind the role of the $U(1)$ 
factors.

\subsection*{Acknowledgment}
We are grateful to Yutaka Yoshida for his early collaboration and also for many valuable discussions 
throughout this work.
M.~H. thanks Harish-Chandra Research Institute and
Niels Bohr Institute for hospitality.
M.~H. is supported by Grant-in-Aid for JSPS fellows (No.22-2764).

\appendix
\section{partition function of $SU(2)_k \times SU(2)_{-k}$ ABJM theory}
\label{app:partition}
Here we provide a further evidence for the conjecture (\ref{iso_N2k2}) 
by calculating the partition function of $SU(2)_k \times SU(2)_{-k}$ ABJM theory on $S^3$. The 
partition function of $U(2)_k \times U(2)_{-k}$ ABJM theory has been exactly calculated 
in \cite{Okuyama:2011su}.

In \cite{Kapustin:2009kz},  the localization technique was applied to the ABJM theory on $S^3$ and its partition function 
was shown to be reduced to a matrix integral
\begin{eqnarray}
Z(N,k)
&=&
\frac{1}{(N!)^2}\int\frac{d^N\mu}{(2\pi)^N}\frac{d^N\nu}{(2\pi)^N} 
\nonumber \\
&~& \quad \quad \quad
\times \frac{\prod_{i<j}\Bigl(2\sinh\frac{\mu_i-\mu_j}{2}\Bigr)^2 
\Bigl(2\sinh\frac{\nu_i-\nu_j}{2}\Bigr)^2}
{\prod_{i,j} \Bigl(2\cosh\frac{\mu_i-\nu_i}{2}\Bigr)^2} 
\exp\left[\frac{ik}{4\pi}\sum_{i=1}^N (\mu_i^2-\nu_i^2)\right], 
\end{eqnarray}
which is commonly referred to as the ABJM matrix model.
Here we consider the constraints
\begin{align}
\mu_1+\mu_2 =0, \quad \nu_1+\nu_2 =0,
\end{align}
for picking up the $SU(2)_k \times SU(2)_{-k}$ factor from 
the $U(2)_k \times U(2)_{-k}$ ABJM matrix model.
Then the partition function of the $SU(2)_k \times SU(2)_{-k}$ theory is given by
\begin{\eqa}
&& Z_{SU(2)_k \times SU(2)_{-k}} \NN\\
&=& \frac{1}{2!}\sum_{\sigma}(-1)^{\sigma}
\int \frac{d^2 \mu }{(2\pi)^2}\frac{d^2 \nu }{(2\pi)^2} 
\prod_i \frac{\exp{\left[ \frac{ik}{4\pi}\sum_{i=1}^2 (\mu^2_i-\nu^2_i) \right]}
            \left[ 2\pi \delta{(\mu_1+\mu_2)} \right] \left[ 2\pi \delta{(\nu_1+\nu_2)} \right] }
                {\Big[ 2\cosh{\left( \frac{\mu_i-\nu_i}{2} \right)} \Big] 
             \left[ 2\cosh{ \left( \frac{\mu_i-\nu_{\sigma(i)}}{2} \right)} \right]} \nonumber \\
&=& \frac{1}{32}
\int \frac{d \mu }{2\pi}\frac{d \nu }{2\pi}
\left[ \frac{1}{\cosh^4{\left( \frac{\mu-\nu}{2} \right)}}-
\frac{1}{\cosh^2{\left( \frac{\mu-\nu}{2} \right)}\cosh^2{\left( \frac{\mu+\nu}{2} \right)}} \right]
\exp{\left[ \frac{ik}{2\pi}(\mu^2-\nu^2) \right]}.
\end{\eqa}
After taking a change of variables
\begin{align}
\lambda=\frac{\mu-\nu}{2}, \quad \lambda'=\frac{\mu+\nu}{2},
\end{align}
we finally obtain
\begin{align}
Z_{SU(2)_k \times SU(2)_{-k}}
&=\frac{1}{32}
\int 2 \frac{d \lambda }{2\pi}\frac{d \lambda' }{2\pi}
\left[ \frac{1}{\cosh^4{\lambda}}-
\frac{1}{\cosh^2{\lambda}\cosh^2{\lambda'}} \right]
\exp{\left[ \frac{2ik}{\pi}\lambda\lambda' \right]} \nonumber \\
&=\frac{1}{64k}-\frac{k}{64}\int d \lambda \frac{2\lambda}{\cosh^2{\pi\lambda}
\sinh{\pi k \lambda}} \nonumber \\
&=\frac{k}{32}\int_{-\infty}^{\infty} d \lambda \frac{\lambda}{\sinh{\pi k \lambda}}\tanh^2{\pi \lambda} \nonumber \\
&=\frac{k}{2}Z_{U(2)_k \times U(2)_{-k}}.
\end{align}
Particularly for $k=2$, 
we find
\begin{\eq}
Z_{U(2)_2 \times U(2)_{-2}}=Z_{SU(2)_2 \times SU(2)_{-2}} .
\end{\eq}
This is consistent with  the expected isomorphism 
between the $U(2)_2 \times U(2)_{-2}$ ABJM theory and  $SU(2)_2 \times SU(2)_{-2}$ BLG theory.

\section{Full result}
Here we show our results for the superconformal indices of various M2-brane theories.
\subsection{$U(2)\times U(2)$ ABJM theory}
\label{sec:ABJMN2}
\begin{longtable}{|l|l|}
  \hline
GNO charges & Index contribution\\
  \hline 
 $T=0$ & $1+4x+12x^2+8x^3+37x^4$ \\
 \hline
 $\ket{0,0}\ket{0,0}$ & $1+4x+12x^2+8x^3+12x^4$\\
 $\ket{1,-1}\ket{1,-1}$ & $25x^4$\\  
  \hline 
  $T=1$ & $5x^2 +24x^3 +23x^4$\\ 
  \hline
$\ket{1,0}\ket{1,0}$ & $5x^2 +24x^3 +23x^4$\\
\hline
$T=2$ & $17x^3+48x^4$\\
\hline
$\ket{1,1}\ket{1,1}$ & $10x^3+16x^4$\\
$\ket{2,0}\ket{2,0}$ & $7x^3+32x^4$\\
\hline
total & $1+4x+8x^{\frac{3}{2}}+12x^2+40x^{\frac{5}{2}}+58x^3+44x^{\frac{7}{2}}+140x^4$\\
\hline
\caption{$U(2)_4\times U(2)_{-4}$.}
\label{ABJM_N2k4}
\end{longtable}

\begin{longtable}{|l|l|}
  \hline
GNO charges & Index contribution\\
  \hline 
 $T=0$ & $1+4x+12x^2+8x^3+12x^4$ \\
 \hline
 $\ket{0,0}\ket{0,0}$ & $1+4x+12x^2+8x^3+12x^4$\\  
  \hline 
  $T=1$ & $6x^{\frac{5}{2}}+28x^{\frac{7}{2}}$\\ 
  \hline
$\ket{1,0}\ket{1,0}$ & $6x^{\frac{5}{2}}+28x^{\frac{7}{2}}$\\
\hline
\caption{$U(2)_5\times U(2)_{-5}$.}
\label{ABJM_N2k5}
\end{longtable}

\begin{longtable}{|l|l|}
  \hline
GNO charges & Index contribution\\
  \hline 
 $T=0$ & $1+4x+12x^2+8x^3+12x^4$ \\
 \hline
 $\ket{0,0}\ket{0,0}$ & $1+4x+12x^2+8x^3+12x^4$\\  
  \hline 
  $T=1$ & $7x^2 +32x^4$\\ 
  \hline
$\ket{1,0}\ket{1,0}$ & $7x^2 +32x^4$\\
\hline
\caption{$U(2)_6\times U(2)_{-6}$.}
\label{ABJM_N2k6}
\end{longtable}

\begin{longtable}{|l|l|}
  \hline
GNO charges & Index contribution\\
  \hline 
 $T=0$ & $1+4x+12x^2+8x^3+12x^4$ \\
 \hline
 $\ket{0,0}\ket{0,0}$ & $1+4x+12x^2+8x^3+12x^4$\\  
  \hline 
  $T=1$ & $8x^{\frac{7}{2}}$\\ 
  \hline
$\ket{1,0}\ket{1,0}$ & $8x^{\frac{7}{2}}$\\
\hline
\caption{$U(2)_7\times U(2)_{-7}$.}
\label{ABJM_N2k7}
\end{longtable}

\subsection{$(SU(2)\times SU(2)) /\mathbb{Z}_2$ BLG theory}
\label{sec:BLG}
\begin{longtable}{|l|l|}
\hline
  GNO charges & Index contribution\\
  \hline 
  $\ket{0,0}\ket{0,0}$ & $1+4x+12x^2+8x^3+12x^4+$\\
 & $z^2(3x+8x^2+12x^3+8x^4)+z^{-2}(3x+8x^2+12x^3+8x^4)+$\\
    & $z^4(6x^2+12x^3+12x^4)+z^{-4}(6x^2+12x^3+12x^4)+$\\
   & $z^6(10x^3+16x^4)+z^{-6}(10x^3+16x^4)+15z^4x^4+15z^{-8}x^4$\\
\hline
 $\ket{1/2,-1/2}\ket{1/2,-1/2}$ & $z(12x^{\frac52}+34x^{\frac72})+z^{-1}(12x^{\frac52}+34x^{\frac72})+$\\  
  & $z^3(10x^{\frac52}+32x^{\frac72})+z^{-3}(10x^{\frac52}+32x^{\frac72})+$\\
  & $z^5(6x^{\frac52}+28x^{\frac72})+z^{-5}(6x^{\frac52}+28x^{\frac72})+14z^7x^{\frac72}+14z^{-7}x^{\frac72}$\\
\hline
\caption{$(SU(2)_5\times SU(2)_{-5})/{\mathbb{Z}}_2$}
\label{BLGZ2_k5}
\end{longtable}

\begin{longtable}{|l|l|}
\hline
  GNO charges & Index contribution\\
  \hline 
  $\ket{0}\ket{0}$ & $1+4x+12x^2+8x^3+12x^4+$\\
 & $z^2(3x+8x^2+12x^3+8x^4)+z^{-2}(3x+8x^2+12x^3+8x^4)+$\\
    & $z^4(6x^2+12x^3+12x^4)+z^{-4}(6x^2+12x^3+12x^4)+$\\
   & $z^6(10x^3+16x^4)+z^{-6}(10x^3+16x^4)+15z^4x^4+15z^{-8}x^4$\\
\hline
 $\ket{1/2}\ket{1/2}$ & $16x^3+41x^4+z^2(15x^3+40x^4)+z^{-2}(15x^3+40x^4)$\\
  & $z^4(12x^3+37x^4)+z^{-4}(12x^3+37x^4)+$\\
  & $z^6(7x^3+32x^4)+z^{-6}(7x^3+32x^4)+16z^8x^4+16z^{-8}x^4$\\
\hline
\caption{$(SU(2)_6\times SU(2)_{-6})/{\mathbb{Z}}_2$}
\label{BLGZ2_k6}
\end{longtable}

\begin{longtable}{|l|l|}
\hline
  GNO charges & Index contribution\\
  \hline 
  $\ket{0}\ket{0}$ & $1+4x+12x^2+8x^3+12x^4+$\\
 & $z^2(3x+8x^2+12x^3+8x^4)+z^{-2}(3x+8x^2+12x^3+8x^4)+$\\
    & $z^4(6x^2+12x^3+12x^4)+z^{-4}(6x^2+12x^3+12x^4)+$\\
   & $z^6(10x^3+16x^4)+z^{-6}(10x^3+16x^4)+15z^4x^4+15z^{-8}x^4$\\
\hline
 $\ket{1/2}\ket{1/2}$ & $20zx^{\frac72}+20z^{-1}x^{\frac72}+18z^3x^{\frac72}+18z^{-3}x^{\frac72}+$\\
  & $14z^5x^{\frac72}+14z^{-5}x^{\frac72}+8z^7x^{\frac72}+8z^{-7}x^{\frac72}$\\
\hline
\caption{$(SU(2)_7\times SU(2)_{-7})/{\mathbb{Z}}_2$}
\label{BLGZ2_k7}
\end{longtable}

\begin{longtable}{|l|l|}
\hline
  GNO charges & Index contribution\\
  \hline 
  $\ket{0}\ket{0}$ & 
$1+ 4x  +12x^2 +8x^3 +12x^4  $\\
 &   $+(3x +8x^2 +12x^3 +8x^4 )z^2 +(3x +8x^2 +12x^3 +8x^4 )z^{-2}$ \\
 &  $+(6x^2 +12x^3 +12x^4 )z^4 +(6x^2 +12x^3 +12x^4 )z^{-4} $ \\ 
\hline
\caption{$(SU(2)_8\times SU(2)_{-8})/\bbZ_2$}
\label{BLG_k8}
\end{longtable}

\begin{longtable}{|l|l|}
\hline
  GNO charges & Index contribution\\
  \hline 
  $\ket{0}\ket{0}$ & 
$1 + 4 x + 12 x^2 + 8 x^3 + 12 x^4 $ \\
 &  $+(3x +8x^2 +12x^3 +8x^4 )z^2 +(3x +8x^2 +12x^3 +8x^4 )z^{-2} $ \\
 &  $+(6x^2 +12x^3 +12x^4 )z^4 +(6x^2 +12x^3 +12x^4 )z^{-4} $ \\
\hline
\caption{$(SU(2)_9\times SU(2)_{-9})/{\mathbb{Z}}_2$}
\label{BLG_Z2k9}
\end{longtable}

\begin{longtable}{|l|l|}
\hline
  GNO charges & Index contribution\\
  \hline 
  $\ket{0}\ket{0}$ & 
$1+ 4x  +12x^2 +8x^3 +12x^4 +15x^4 z^8 +15 x^4 z^{-8}$ \\
&  $ +\left(16 x^4+10 x^3 \right) z^6 +(16 x^4+10 x^3)z^{-6}$\\
&$+\left(12 x^4+12 x^3+6 x^2\right) z^4 +(12 x^4+12 x^3+6x^2)z^{-4}$\\
&$+\left(8 x^4+12 x^3+8 x^2+3 x\right) z^2 +(8 x^4+12 x^3+8 x^2+3 x)z^{-2}$\\
\hline
\caption{$(SU(2)_{16}\times SU(2)_{-16})/{\mathbb{Z}}_2$}
\label{BLG_Z2k16}
\end{longtable}

\subsection{$U(3)\times U(3)$ ABJM theory}
\label{sec:ABJM_N3}
\begin{longtable}{|l|l|}
  \hline
GNO charges & Index contribution\\
  \hline 
 $T=0$ & $1+8x+71x^2+320x^3$ \\
 \hline
 $\ket{0,0,0}\ket{0,0,0}$ & $1+4x+12x^2+32x^3$\\  
 \hline
 $\ket{1,0,-1}\ket{1,0,-1}$ & $4x+32x^2+92x^3 $\\  
 \hline
 $\ket{2,0,-2}\ket{2,0,-2}$ & $9x^2+60x^3 $\\  
 \hline
 $\ket{1,1,-2}\ket{1,1,-2}$ & $9x^2+36x^3 $\\  
 \hline
 $\ket{-1,-1,2}\ket{-1,-1,2}$ & $9x^2+36x^3 $\\  
 \hline
 $\ket{1,2,-3}\ket{1,2,-3}$ & $24x^3 $\\  
 \hline
 $\ket{-1,-2,3}\ket{-1,-2,3}$ & $24x^3 $\\  
 \hline
 $\ket{3,0,-3}\ket{3,0,-3}$ & $16x^3 $\\  
  \hline 
  $T=1$ & $2x^{1/2}+24x^{3/2}+156x^{5/2}$\\ 
  \hline
$\ket{1,0,0}\ket{1,0,0}$ & $2x^{1/2}+12x^{3/2}+42x^{5/2}$\\
  \hline
$\ket{1,1,-1}\ket{1,1,-1}$ & $6x^{3/2}+28x^{5/2}$\\
  \hline
$\ket{2,0,-1}\ket{2,0,-1}$ & $6x^{3/2}+44x^{5/2}$\\
  \hline
$\ket{2,1,-2}\ket{2,1,-2}$ & $18x^{5/2}$\\
  \hline
$\ket{3,0,-2}\ket{3,0,-2}$ & $12x^{5/2}$\\
  \hline
$\ket{3,-1,-1}\ket{3,-1,-1}$ & $12x^{5/2}$\\
  \hline 
  $T=2$ & $6x+56x^2 +311x^3 $\\ 
  \hline
$\ket{2,0,0}\ket{2,0,0}$ & $3x+16x^2 +52x^3 $\\
  \hline
$\ket{1,1,0}\ket{1,1,0}$ & $3x+20x^2 +51x^3 $\\
  \hline
$\ket{2,1,-1}\ket{2,1,-1}$ & $12x^2 +64x^3 $\\
  \hline
$\ket{3,0,-1}\ket{3,0,-1}$ & $8x^2 +56x^3 $\\
  \hline
$\ket{3,1,-2}\ket{3,1,-2}$ & $24x^3 $\\
  \hline
$\ket{2,2,-2}\ket{2,2,-2}$ & $18x^3 $\\
  \hline
$\ket{4,0,-2}\ket{4,0,-2}$ & $15x^3 $\\
  \hline
$\ket{4,-1,-1}\ket{4,-1,-1}$ & $15x^3 $\\
\hline
  $T=3$ & $14x^{3/2}+114x^{5/2} $\\ 
  \hline
$\ket{3,0,0}\ket{3,0,0}$ & $4x^{3/2}+20x^{5/2} $\\
  \hline
$\ket{2,1,0}\ket{2,1,0}$ & $6x^{3/2}+44x^{5/2} $\\
  \hline
$\ket{1,1,1}\ket{1,1,1}$ & $4x^{3/2}+12x^{5/2} $\\
  \hline
$\ket{3,1,-1}\ket{3,1,-1}$ & $16x^{5/2} $\\
  \hline
$\ket{2,2,-1}\ket{2,2,-1}$ & $12x^{5/2} $\\
  \hline
$\ket{4,0,-1}\ket{4,0,-1}$ & $10x^{5/2}$\\
\hline
\caption{$U(3)_1\times U(3)_{-1}$.}
\label{ABJM_N3k1}
\end{longtable}

\begin{longtable}{|l|l|}
  \hline
GNO charges & Index contribution\\
  \hline 
 $T=0$ & $1+4x+21x^2+92x^3$ \\
 \hline
 $\ket{0,0,0}\ket{0,0,0}$ & $1+4x+12x^2+32x^3$\\  
 \hline
 $\ket{1,0,-1}\ket{1,0,-1}$ & $9x^2+60x^3 $\\  
  \hline 
  $T=1$ & $3x +16x^2 +87x^3 $\\ 
  \hline
$\ket{1,0,0}\ket{1,0,0}$ & $3x +16x^2 +54x^3$\\
  \hline
$\ket{2,0,-1}\ket{2,0,-1}$ & $15x^3 $\\
  \hline
$\ket{1,1,-1}\ket{1,1,-1}$ & $18x^3 $\\
  \hline 
  $T=2$ & $11x^2 +60x^3 $\\ 
  \hline
$\ket{2,0,0}\ket{2,0,0}$ & $5x^2 +24x^3 $\\
  \hline
$\ket{1,1,0}\ket{1,1,0}$ & $6x^2 +36x^3 $\\
\hline
\caption{$U(3)_2\times U(3)_{-2}$.}
\label{ABJM_N3k2}
\end{longtable}

\begin{longtable}{|l|l|}
  \hline
GNO charges & Index contribution\\
  \hline 
 $T=0$ & $1+4x+12x^2+48x^3$ \\
 \hline
 $\ket{0,0,0}\ket{0,0,0}$ & $1+4x+12x^2+32x^3$\\  
 \hline
 $\ket{1,0,-1}\ket{1,0,-1}$ & $16x^3 $\\  
  \hline 
  $T=1$ & $4x^{3/2} +20x^{5/2} $\\ 
  \hline
$\ket{1,0,0}\ket{1,0,0}$ & $4x^{3/2} +20x^{5/2} $\\
  \hline 
  $T=2$ & $17x^3$\\ 
  \hline
$\ket{2,0,0}\ket{2,0,0}$ & $7x^3 $\\
  \hline
$\ket{1,1,0}\ket{1,1,0}$ & $10x^3 $\\
\hline 
\caption{$U(3)_3\times U(3)_{-3}$.}
\label{ABJM_N3k3}
\end{longtable}

\begin{longtable}{|l|l|}
  \hline
GNO charges & Index contribution\\
  \hline 
 $T=0$ & $1+4x+12x^2+32x^3$ \\
 \hline
 $\ket{0,0,0}\ket{0,0,0}$ & $1+4x+12x^2+32x^3$\\  
  \hline 
  $T=1$ & $5x^2 +24x^3 $\\ 
  \hline
$\ket{1,0,0}\ket{1,0,0}$ & $5x^2 +24x^3 $\\
\hline
\caption{$U(3)_4\times U(3)_{-4}$.}
\label{ABJM_N3k4}
\end{longtable}

\begin{longtable}{|l|l|}
  \hline
GNO charges & Index contribution\\
  \hline 
 $T=0$ & $1+4x+12x^2+32x^3$ \\
 \hline
 $\ket{0,0,0}\ket{0,0,0}$ & $1+4x+12x^2+32x^3$\\  
  \hline 
  $T=1$ & $6x^{5/2} $\\ 
  \hline
$\ket{1,0,0}\ket{1,0,0}$ & $6x^{5/2} $\\
  \hline 
\caption{$U(3)_5\times U(3)_{-5}$.}
\label{ABJM_N3k5}
\end{longtable}

\subsection{$(SU(3)\times SU(3))/\bbZ_3$ theory}
\label{sec:SU3}

\begin{longtable}{|l|l|}
\hline
  GNO charges & Index contribution\\
  \hline 
  $\ket{0,0,0}\ket{0,0,0}$ & $1+4x+12x^2+32x^3$\\
    & $z^3 (4x^{3/2} +12x^{5/2}) +z^{-3} (4x^{3/2} +12x^{5/2})$\\
   & $10z^6x^3+10z^{-6}x^3$\\
\hline
$\ket{\frac{1}{3},\frac{1}{3},-\frac{2}{3}}\ket{\frac{1}{3},\frac{1}{3},-\frac{2}{3}}$ 
& $12z^{-5} x^{5/2}+z \left(42 x^{5/2}+12 x^{3/2}+2 \sqrt{x}\right)$ \\
  &      $+z^4 \left(36 x^3+9 x^2\right)+z^{-2}(51 x^3+20 x^2+3 x)$\\
\hline
$\ket{-\frac{1}{3},-\frac{1}{3},\frac{2}{3}}\ket{-\frac{1}{3},-\frac{1}{3},\frac{2}{3}}$ 
& $12z^5 x^{5/2}+z^{-1}(42 x^{5/2}+12 x^{3/2}+2 \sqrt{x})$ \\
  &      $+z^{-4}(36 x^3+9 x^2)+z^2 \left(51 x^3+20 x^2+3 x \right)$\\
\hline
$\ket{\frac{2}{3},\frac{2}{3},-\frac{4}{3}}\ket{\frac{2}{3},\frac{2}{3},-\frac{4}{3}}$ 
& $2z^5 x^{5/2}+z^{-1}(28 x^{5/2}+6 x^{3/2})+z^{-4}(36 x^3+6 x^2)+z^2\left(52 x^3+16 x^2+3 x \right)$ \\
\hline
$\ket{-\frac{2}{3},-\frac{2}{3},\frac{4}{3}}\ket{-\frac{2}{3},-\frac{2}{3},\frac{4}{3}}$ 
& $12z^5 x^{5/2}+z\left(28 x^{5/2}+6 x^{3/2}\right)+z^4\left(36 x^3+6 x^2\right)+z^{-2}(52 x^3+16 x^2+3 x)$ \\
\hline
$\ket{1,0,-1}\ket{1,0,-1}$ & $4x+32x^2 +92x^3$\\
 & $+z^3( 6x^{3/2}+44x^{5/2})+z^{-3}( 6x^{3/2}+44x^{5/2}) +24z^6x^3+24z^{-6}x^3$\\
\hline
$\ket{1,1,-2}\ket{1,1,-2}$ & $9x^2 +36x^3$\\
 & $+z^3( 4x^{3/2}+20x^{5/2})+12z^{-3}x^{5/2} +10z^{-6}x^3$\\
\hline
$\ket{-1,-1,2}\ket{-1,-1,2}$ & $9x^2 +36x^3$\\
 & $+z^3 (12x^{5/2}) +z^{-3}( 4x^{3/2}+20x^{5/2}) +10z^{-6}x^3$\\
\hline
$\ket{\frac{1}{3},\frac{4}{3},-\frac{5}{3}}\ket{\frac{1}{3},\frac{4}{3},-\frac{5}{3}}$ 
& $12z^{-5} x^{5/2}+z\left(44 x^{5/2}+6 x^{3/2}\right)+z^4\left(56 x^3+8 x^2\right)+z^{-2}(64 x^3+12 x^2)$\\
\hline
$\ket{-\frac{1}{3},-\frac{4}{3},\frac{5}{3}}\ket{-\frac{1}{3},-\frac{4}{3},\frac{5}{3}}$ 
& $ 12 z^5x^{5/2}+z^{-1}(44 x^{5/2}+6 x^{3/2})+z^{-4}(56 x^3+8 x^2)+z^2\left(64 x^3+12 x^2\right)$\\
\hline
$\ket{\frac{2}{3},\frac{5}{3},-\frac{7}{3}}\ket{\frac{2}{3},\frac{5}{3},-\frac{7}{3}}$ 
& $ 10z^5 x^{5/2}+18z^{-1}x^{5/2}+24z^{-4} x^3+z^2\left(56 x^3+8 x^2\right)$\\
\hline
$\ket{-\frac{2}{3},-\frac{5}{3},\frac{7}{3}}\ket{-\frac{2}{3},-\frac{5}{3},\frac{7}{3}}$ 
& $ 10z^{-5} x^{5/2}+18z x^{5/2} +24z^4 x^3 +z^{-2}\left(56 x^3+8 x^2\right)$\\
\hline
$\ket{\frac{4}{3},\frac{4}{3},-\frac{8}{3}}\ket{\frac{4}{3},\frac{4}{3},-\frac{8}{3}}$ 
& $ 12z x^{5/2}+18z^{-2} x^3+z^4\left(24 x^3+5 x^2\right)$\\
\hline
$\ket{-\frac{4}{3},-\frac{4}{3},\frac{8}{3}}\ket{-\frac{4}{3},-\frac{4}{3},\frac{8}{3}}$ 
& $ 12z^{-1} x^{5/2}+18z^2 x^3+z^{-4}\left(24 x^3+5 x^2\right)$\\
\hline
$\ket{\frac{5}{3},\frac{5}{3},-\frac{10}{3}}\ket{\frac{5}{3},\frac{5}{3},-\frac{10}{3}}$ & $6 z^5x^{5/2}+15z^2 x^3$\\
\hline
$\ket{2,0,-2}\ket{2,0,-2}$ & $9x^2 +60x^3$\\
 & $+16z^3x^{5/2}+16z^{-3}16x^{5/2} +15z^6x^3+15z^{-6}x^3$\\
\hline
$\ket{1,2,-3}\ket{1,2,-3}$ & $24x^3 +10z^3x^{5/2}+12z^6x^3$\\
\hline
$\ket{-1,-2,3}\ket{-1,-2,3}$ & $24x^3 +10z^{-3}x^{5/2} +12z^{-6}x^3$\\
\hline
$\ket{\frac{1}{3},\frac{7}{3},-\frac{8}{3}}\ket{\frac{1}{3},\frac{7}{3},-\frac{8}{3}}$ 
& $12 zx^{5/2}+20z^4 x^3+24z^{-2} x^3$\\
\hline
$\ket{\frac{2}{3},\frac{8}{3},-\frac{10}{3}}\ket{\frac{2}{3},\frac{8}{3},-\frac{10}{3}}$ & $15z^2 x^3$\\
\hline
$\ket{\frac{4}{3},\frac{7}{3},-\frac{11}{3}}\ket{\frac{4}{3},\frac{7}{3},-\frac{11}{3}}$   &   $12z^4 x^3$\\
\hline
$\ket{3,0,-3}\ket{3,0,-3}$ & $16x^3$\\
\hline
$\ket{4,-2,-2}\ket{4,-2,-2}$ & $7z^{-6}x^3$\\
\hline
$\ket{-4,2,2}\ket{-4,2,2}$ & $7z^6x^3$\\
\hline
Total & $1+8x+71x^2 +320x^3 +(2x^{1/2} +24x^{3/2} +156x^{5/2})z  $  \\
      & $+(6x +56x^2 +293x^3 ) z^2 +(14x^{3/2} +114x^{5/2})z^3$ \\
\hline
\caption{$(SU(3)_1 \times SU(3)_{-1})/{\mathbb{Z}}_3$}
\label{SU3_k1}
\end{longtable}

\begin{longtable}{|l|l|}
\hline
  GNO charges & Index contribution\\
  \hline 
  $\ket{0,0,0}\ket{0,0,0}$ & 
$1+4 x+12 x^2+32 x^3+z^3(4x^{3/2}+12 x^{5/2})$ \\
 & $z^{-3}(4x^{3/2}+12 x^{5/2})+10z^6 x^3+10z^{-6} x^3$\\
\hline
$\ket{\frac{1}{3},\frac{1}{3},-\frac{2}{3}} \ket{\frac{1}{3},\frac{1}{3},-\frac{2}{3}}$ & 
$12z^5 x^{5/2}+z^{-1}(30 x^{5/2}+6 x^{3/2})$ \\
& $+z^{-4}(36 x^3+6 \ x^2)+z^2\left(54 x^3+16 x^2+3 x\right)$\\
\hline
$\ket{-\frac{1}{3},-\frac{1}{3},\frac{2}{3}} \ket{-\frac{1}{3},-\frac{1}{3},\frac{2}{3}}$ & 
$12z^{-5} x^{5/2}+z\left(30 x^{5/2}+6 x^{3/2}\right)$ \\
& $+z^4\left(36 \ x^3+6 x^2\right)+z^{-2}(54 x^3+16 x^2+3 x)$\\
\hline
$\ket{\frac{2}{3},\frac{2}{3},-\frac{4}{3}} \ket{\frac{2}{3},\frac{2}{3},-\frac{4}{3}}$ & 
$ 12z x^{5/2}+18z^{-2} x^3+z^4\left(24 x^3+5 x^2\right)$\\
\hline
$\ket{-\frac{2}{3},-\frac{2}{3},\frac{4}{3}} \ket{-\frac{2}{3},-\frac{2}{3},\frac{4}{3}}$ & 
$ 12z^{-1} x^{5/2}+18z^2 x^3+z^{-4}(24 x^3+5 x^2)$\\
\hline
$\ket{1,0,-1}\ket{1,0,-1}$ & 
$16z^3 x^{5/2}+16z^{-3} x^{5/2}+15z^6 x^3+15z^{-6}x^3+60 x^3+9 x^2 $\\
\hline
$\ket{\frac{1}{3},\frac{4}{3},-\frac{5}{3}} \ket{\frac{1}{3},\frac{4}{3},-\frac{5}{3}}$ & $15z^2 x^3$\\
\hline
$\ket{-\frac{1}{3},-\frac{4}{3},\frac{5}{3}} \ket{-\frac{1}{3},-\frac{4}{3},\frac{5}{3}}$ & $15z^{-2} x^3$\\
\hline
$\ket{1,1,-2}\ket{1,1,-2}$ & $7z^{6}x^3$ \\
\hline
$\ket{-1,-1,2}\ket{-1,-1,2}$  & $7z^{-6}x^3$\\
\hline
Total & $1+4x+21x^2 +92x^3 +(3x +16x^2 +87x^3 )z^2 $\\
        & $+(11x^2 +60x^3 )z^4$ \\
\hline
\caption{$(SU(3)_2 \times SU(3)_{-2})/{\mathbb{Z}}_3$}
\label{SU3_k2}
\end{longtable}

\begin{longtable}{|l|l|}
\hline
  GNO charges & Index contribution\\
  \hline 
  $\ket{0,0,0}\ket{0,0,0}$ & $1+4x+12x^2+32x^3$\\
   & $+z^3 (4x^{3/2} +12x^{5/2}) +z^{-3} (4x^{3/2} +12x^{5/2})$\\
   & $+10z^6x^3+10z^{-6}x^3$\\
\hline
$\ket{1,0,-1}\ket{1,0,-1}$ & $16x^3$\\
\hline
$\ket{1/3,1/3,-2/3}\ket{1/3,1/3,-2/3}$ 
& $9x^2 +40x^3 +z^3 (4x^{3/2}+20x^{5/2}) +z^{-3}(12x^{5/2})$\\
& $+15z^6x^3 +10z^{-6}x^3$ \\
\hline
$\ket{-1/3,-1/3,2/3}\ket{-1/3,-1/3,2/3}$ 
& $9x^2 +40x^3 +z^3 (12x^{5/2}) +z^{-3}(4x^{3/2}+20x^{5/2})$\\
& $+10z^6x^3+15z^{-6}x^3$ \\
\hline
$\ket{4/3,-2/3,-2/3}\ket{4/3,-2/3,-2/3}$ & $7z^{-6}x^3$\\
\hline
$\ket{-4/3,2/3,2/3}\ket{-4/3,2/3,2/3}$ & $7z^{6}x^3$\\
\hline
Total & $1+4x+30x^2 +128x^3 +z^3(8x^{3/2} +32x^{5/2})+42z^6x^3$ \\
\hline
\caption{$(SU(3)_3 \times SU(3)_{-3})/{\mathbb{Z}}_3$}
\label{SU3_k3}
\end{longtable}

\begin{longtable}{|l|l|}
\hline
  GNO charges & Index contribution\\
  \hline 
  $\ket{0,0,0}\ket{0,0,0}$ & $1+4x+12x^2+32x^3$\\
   & $+z^3 (4x^{3/2} +12x^{5/2}) +z^{-3} (4x^{3/2} +12x^{5/2})$\\
   & $+10z^6x^3+10z^{-6}x^3$\\
  \hline 
  $\ket{\frac{1}{3},\frac{1}{3},-\frac{2}{3}}\ket{\frac{1}{3},\frac{1}{3},-\frac{2}{3}}$ 
& $12z x^{5/2}+18z^{-2} x^3+z^4\left(24 x^3+5 x^2\right)$\\
  \hline 
  $\ket{-\frac{1}{3},-\frac{1}{3},\frac{2}{3}}\ket{-\frac{1}{3},-\frac{1}{3},\frac{2}{3}}$ 
& $12z^{-1} x^{5/2}+18z^2 x^3+z^{-4}(24 x^3+5 x^2)$\\
\hline
Total & $1+4x +12x^2 +32x^3 +(5x^2 +24x^3 )z^4$\\
\hline
\caption{$(SU(3)_4 \times SU(3)_{-4})/{\mathbb{Z}}_3$}
\label{SU3_k4}
\end{longtable}

\begin{longtable}{|l|l|}
\hline
  GNO charges & Index contribution\\
  \hline 
  $\ket{0,0,0}\ket{0,0,0}$ & $1+4x+12x^2+32x^3$\\
   & $+z^3 (4x^{3/2} +12x^{5/2}) +z^{-3} (4x^{3/2} +12x^{5/2})$\\
   & $+10z^6x^3+10z^{-6}x^3$\\
  \hline 
  $\ket{\frac{1}{3},\frac{1}{3},-\frac{2}{3}}\ket{\frac{1}{3},\frac{1}{3},-\frac{2}{3}}$ 
& $6z^5 x^{5/2}+15z^2 x^3$\\
  \hline 
  $\ket{-\frac{1}{3},-\frac{1}{3},\frac{2}{3}}\ket{-\frac{1}{3},-\frac{1}{3},\frac{2}{3}}$ 
& $6z^{-5} x^{5/2}+15z^{-2} x^3$\\
\hline
Total & $1+4x +12x^2 +32x^3 +6 x^{5/2} z^5 $\\
\hline
\caption{$(SU(3)_5 \times SU(3)_{-5})/{\mathbb{Z}}_3$}
\label{SU3_k5}
\end{longtable}

\begin{longtable}{|l|l|}
\hline
  GNO charges & Index contribution\\
  \hline 
  $\ket{0,0,0}\ket{0,0,0}$ & $1+4x+12x^2+32x^3$\\
   & $+z^3 (4x^{3/2} +12x^{5/2}) +z^{-3} (4x^{3/2} +12x^{5/2})$\\
   & $+10z^6x^3+10z^{-6}x^3$\\
\hline
$\ket{1/3,1/3,-2/3}\ket{1/3,1/3,-2/3}$ & $7z^6x^3$ \\
\hline
$\ket{-1/3,-1/3,2/3}\ket{-1/3,-1/3,2/3}$ & $7z^{-6}x^3$ \\
\hline
\caption{$(SU(3)_6 \times SU(3)_{-6})/{\mathbb{Z}}_3$}
\label{SU3_k6}
\end{longtable}

\subsection{$U(3)\times U(2)$ ABJ theory}
\label{sec:ABJ_N32}

\begin{longtable}{|l|l|}
  \hline
GNO charges & Index contribution\\
  \hline 
 $T=0$ & $1+4x+12x^2+28x^3+37x^4$ \\
 \hline
 $\ket{0,0,0}\ket{0,0}$ & $1+4x+12x^2+12x^3+5x^4$\\  
 \hline
 $\ket{1,0,-1}\ket{1,-1}$ & $16x^3 +32x^4$\\  
  \hline 
  $T=1$ & $4x^{\frac{3}{2}}+20x^{\frac{5}{2}} +26x^{\frac{7}{2}}$\\ 
  \hline
$\ket{1,0,0}\ket{1,0}$ & $4x^{\frac{3}{2}}+20x^{\frac{5}{2}} +26x^{\frac{7}{2}}$\\
  \hline 
  $T=2$ & $17x^3 +48x^4$\\ 
  \hline
$\ket{2,0,0}\ket{2,0}$ & $7x^3 +32x^4$\\
  \hline
$\ket{1,1,0}\ket{1,1}$ & $10x^3 +16x^4$\\
\hline
\caption{$U(3)_3\times U(2)_{-3}$.}
\label{U32_k3}
\end{longtable}

\begin{longtable}{|l|l|}
  \hline
GNO charges & Index contribution\\
  \hline 
 $T=0$ & $1+4x+12x^2+12x^3+30x^4$ \\
 \hline
 $\ket{0,0,0}\ket{0,0}$ & $1+4x+12x^2+12x^3+5x^4$\\  
 \hline
 $\ket{1,0,-1}\ket{1,-1}$ & $25x^4$\\  
  \hline 
  $T=1$ & $5x^2 +24x^3 +28x^4$\\ 
  \hline
$\ket{1,0,0}\ket{1,0}$ & $5x^2 +24x^3 +28x^4$\\
  \hline 
  $T=2$ & $24x^4$\\ 
  \hline
$\ket{2,0,0}\ket{2,0}$ & $9x^4$\\
  \hline
$\ket{1,1,0}\ket{1,1}$ & $15x^4$\\
\hline
\caption{$U(3)_4\times U(2)_{-4}$.}
  \label{U32_k4}
\end{longtable}

\subsection{$U(4)\times U(2)$ ABJ theory}
\label{sec:ABJ_N42}

\begin{longtable}{|l|l|}
  \hline
GNO charges & Index contribution\\
  \hline 
 $T=0$ & $1+4x+12x^2+12x^3+31x^4$ \\
 \hline
 $\ket{0,0,0,0}\ket{0,0}$ & $1+4x+12x^2+12x^3+6x^4$\\  
 \hline
 $\ket{1,0,0,-1}\ket{1,-1}$ & $25x^4$\\  
  \hline 
  $T=1$ & $5x^2 +24x^3 +28x^4$\\ 
  \hline
$\ket{1,0,0,0}\ket{1,0}$ & $5x^2 +24x^3 +28x^4$\\
  \hline 
\caption{$U(4)_4\times U(2)_{-4}$.}
  \label{U42_k4}
\end{longtable}

\bibliographystyle{JHEP}
\bibliography{index}

\end{document}